\documentclass[journal]{IEEEtran}

\usepackage{lineno}
\modulolinenumbers[5]
\usepackage{color}
\usepackage{graphicx}
\usepackage{balance}  
\usepackage{subfigure}
\usepackage{times, amsmath, epsfig,amssymb,booktabs}
\usepackage{multicol}
\usepackage{multirow}
\usepackage{verbatim}
\usepackage{algorithmic}
\usepackage[ruled,vlined]{algorithm2e}
\usepackage{bm}
\usepackage{multirow}
\usepackage{cite}
\usepackage{epstopdf}
\usepackage{diagbox}
\usepackage[shortcuts]{extdash}

\usepackage{makecell}
\begin{document}

	
	
	\title{Fair Influence Maximization in Social Networks: A Community-Based Evolutionary Algorithm}

\author{Kaicong~Ma, Xinxiang~Xu,
        Haipeng~Yang, Renzhi~Cao, and Lei~Zhang,~\IEEEmembership{Member,~IEEE}
\IEEEcompsocitemizethanks{\IEEEcompsocthanksitem This work was supported by the National Natural Science Foundation of China (61976001, 61876184), and the Key Projects of University Excellent Talents Support Plan of Anhui Provincial Department of Education (gxyqZD2021089). \it{(Corresponding author: Lei~Zhang.)}}

\IEEEcompsocitemizethanks{\IEEEcompsocthanksitem K. Ma, X. Xu, H. Yang, and L. Zhang are with Information Materials and Intelligent Sensing Laboratory of Anhui Province, School of Computer Science and Technology, Anhui University, Hefei 230039, China (email: mkc17@foxmail.com; xinxiangxu22@foxmail.com; haipengyang@126.com; zl@ahu.edu.cn).}

\IEEEcompsocitemizethanks{\IEEEcompsocthanksitem R. Cao is with School of Computer Science, Pacific Lutheran University, Washington State, USA (email: caora@plu.edu).}}


	\maketitle
	
	\begin{abstract}
		
		Influence Maximization (IM) has been extensively studied in network science, which attempts to find a subset of users to maximize the influence spread. A new variant of IM, Fair Influence Maximization (FIM), which primarily enhances the fair propagation of information, attracts increasing attention in academic. However, existing algorithms for FIM suffer from a trade-off between fairness and running time. Since it is a tough task to ensure that users are fairly influenced in terms of sensitive attributes, such as race or gender, while maintaining a high influence spread. To tackle this problem, in this paper, we propose an effective and efficient Community-based Evolutionary Algorithm for FIM (named CEA-FIM). In CEA-FIM, a community-based node selection strategy is proposed to identify potential nodes, which not only considers the size of the community but also the attributes of the nodes in the community. Subsequently, we design an evolutionary algorithm based on the proposed node selection strategy to hasten the search for the optimal solution, including the novel initialization, crossover and mutation strategies. We validate the proposed algorithm CEA-FIM by performing experiments on real-world and synthetic networks. The experimental results show that the proposed CEA-FIM achieves a better balance between effectiveness and efficiency, compared to the state-of-the-art baseline algorithms.
		
	\end{abstract}
	

	\section{Introduction}
	\IEEEPARstart{I}{nfluence} Maximization (IM), which selects a subset of nodes to maximize the number of final influenced nodes, has attracted much attention from scholars~\cite{li2018influence}. The IM problem has been well studied over the past twenty years and has a considerable number of applications, such as HIV prevention for homeless youth~\cite{yadav2018bridging}, job vacancy advertisements~\cite{chen2012time}, rumor control~\cite{he2012influence}, financial inclusion~\cite{banerjee2013diffusion}, and more. Domingos and Richardson were the first to propose the IM problem~\cite{domingos2001mining}. Kempe \emph{et al.}~\cite{kempe2003maximizing} showed that the IM problem is NP-hard and proposed a standard greedy algorithm to tackle it. Thereafter, Leskovec \emph{et al.}~\cite{leskovec2007cost} proposed an improved greedy algorithm known as the ``Cost-Effective Lazy Forward" (CELF) scheme. CELF uses the submodularity of the objective function to reduce the number of Monte Carlo simulations. In addition, there are numerous algorithms in the literature, including the simulation-based approach~\cite{kempe2003maximizing,leskovec2007cost}, the heuristic approach~\cite{kim2013scalable,chen2020scalable} and the metaheuristic approach~\cite{wang2019finding,zhang2022search}.
	
	However, IM only focuses on the spread of influence, while ignoring the disparity in minority groups. In recent years, due to the increasing deployment of automated decision-making systems in people's daily lives, there has been growing concern about their potential discrimination against people, especially those belonging to small and marginalized groups. Such systems, which focus solely on the greatest gain, usually do not take into account the interests of minorities, which can lead to unequal treatment of some groups in real life. Taking a practical application of IM as an example, nearly 2 million homeless young people in the United States are vulnerable to HIV infection~\cite{pfeifer1997study,yadav2018bridging}. Due to the limitation of human and financial resources, it is recommended to select and train only a small number of homeless youth, which enables them to disseminate HIV prevention knowledge to those they interact with. In fact, there are groups of different genders, ages, and races among these homeless youth. It is of cardinal significance to consider these attributes when selecting initial homeless youth, so that individuals in different groups can learn about HIV protection. Consequently, fairness---the elimination of all unequal treatment of individuals or groups---has recently drawn much attention from scholars~\cite{mehrabi2021survey,caton2020fairness,dong2022fairness}. Not surprisingly, a growing body of work attempts to enhance the fairness of IM algorithms~\cite{tsang2019group,khajehnejad2021adversarial,khajehnejad2022crosswalk,razaghi2022group,gong2023influence}.
	
	Fair Influence Maximization (FIM) aims to ensure that individuals or groups receive the social resources they deserve, which are critical to their daily lives. The existing FIM problems focus on individual fairness~\cite{fish2019gaps,stoica2019fairness} and group fairness~\cite{tsang2019group,khajehnejad2021adversarial,khajehnejad2022crosswalk,stoica2019fairness}. While individual fairness considers similar treatment between two similar individuals, group fairness focuses on the equal allocation of resources among different groups. Given the fact that the definition of group fairness is accepted by most of the existing literature, we focus on group fairness which is more realistic than individual fairness. Tsang \emph{et al.}~\cite{tsang2019group} formalized the concept of group fairness in IM and proposed an algorithm (denoted as DC) to find solutions that satisfy fairness. However, the time complexity of DC is quite high, which means that it is difficult to apply it to large-scale social networks. Afterwards, embedding-based methods~\cite{khajehnejad2021adversarial,khajehnejad2022crosswalk} were proposed to solve the FIM problem, which have high scalability. Multi-objective optimization algorithms~\cite{razaghi2022group,gong2023influence} were also employed to obtain some Pareto non-dominated solutions to the FIM problem. \textcolor{black}{In general, existing FIM algorithms are hard to strike a good balance between effectiveness and efficiency, which lose sight of the magnitude of social network characteristics, such as community structure.}
	
	\begin{figure}[!htb]\vspace{-0.5cm}
		\begin{center}
			\centering
			\subfigure{
				\includegraphics[scale=0.4]{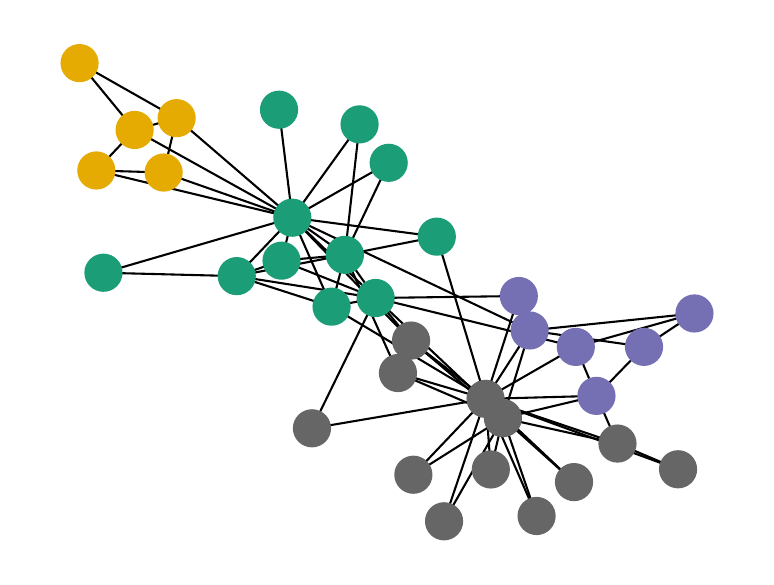}
			}
			\caption{The community structure of Zachary's karate club network~\cite{zachary1977information}, where nodes with the same color belong to the same community.}\vspace{-0.5cm}
			\label{fig:community}
		\end{center}
	\end{figure}\vspace{-0.0cm}
	
	Intuitively, a community is a densely connected subset of nodes that are only sparsely linked to the remaining network~\cite{girvan2002community}. For instance, Fig.~\ref{fig:community} shows the community structure of Zachary's karate club network~\cite{zachary1977information} detected by Louvain algorithm~\cite{blondel2008fast}. Community-based methods~\cite{wang2010community,styczen2022targeted,zhang2021local,li2018community} have been proven to be effective in solving the IM problem and its variants, which can tremendously reduce the prohibitive cost of identifying potential nodes over the whole network. For example, Styczen \emph{et al.}~\cite{styczen2022targeted} utilized the community structure to solve the disparity influence maximization problem, which is the most related to our work. Yet, there is still an urgent need to take advantage of the community structure to efficiently and effectively solve the FIM problem. The main technical challenge is how to exploit the community structure property of social networks to mine nodes that can disseminate information fairly across different groups. It is also worth highlighting that some FIM methods impose fairness as strict constraints, which leads to undesirable properties such as resource waste.

	To this end, we propose a community-based evolutionary algorithm called CEA-FIM for the FIM problem. The proposed algorithm CEA-FIM has good effectiveness and efficiency, which are mainly attributed to the utilization of community structure information. The main contributions of this paper are summarized as follows:
	
	\begin{enumerate}
		\item[1)] \textcolor{black}{We propose a community-based node selection strategy to find potential nodes, which takes into account the size of community and the attributes of the nodes in the community. Using community structure information, we obtain nodes that fairly disseminate information among sensitive attributes. Note that we are the first to exploit community structure information to solve the FIM problem effectively and efficiently.}
	
		\item[2)] \textcolor{black}{We transform the existing FIM problem into a single-objective optimization problem and propose a community-based evolutionary algorithm denoted as CEA-FIM to address it. To be specific, in order to speed up the optimal solution search process, the community-based node selection strategy is fully integrated into CEA-FIM, including the novel suggested initialization, crossover and mutation strategies.}
	
		\item[3)] We conduct extensive experiments on both real-world and synthetic networks to evaluate the performance of the proposed algorithm CEA-FIM. The results validate that the proposed algorithm achieves a better balance in terms of fairness and running time than the state-of-the-art baseline algorithms.
	\end{enumerate}
	
	The rest of this paper is organized as follows. Section~\ref{sec:prel} introduces the preliminaries, including problem formulation, information diffusion model, evaluation function, and related work. Then, the proposed algorithm CEA-FIM including community-based node selection strategy and evolutionary framework is described in Section~\ref{sec:algo}. Next, the experimental results and analysis of the proposed algorithm CEA-FIM are presented in Section~\ref{sec:expe}. Finally, the conclusion and the future work are given in Section~\ref{sec:conclude}.
	
	\section{Preliminaries And Related Work}\label{sec:prel}
	This section first introduces some preliminaries about Fair Influence Maximization (FIM), including problem formulation, information diffusion model and evaluation function. Then, we mainly review related work on community-based IM methods and FIM methods.
	
	\subsection{Preliminaries}
	
	\subsubsection{FIM Problem}
	A real-world social network is modeled as a graph $G(V,E)$. Users in a social network are represented as a node set $V$, and the connection between any two users is represented as an edge set $E$. Let $n$ be the number of nodes in $G$, i.e., $n=|V|$. Assume that all nodes in $G$ can be divided into $q$ groups, i.e., $R=\{R_1,R_2,...,R_q\}$ where each node belongs to \textcolor{black}{one or more group}, i.e., $V_{R_1}\cup V_{R_2}\cup ...\cup V_{R_q}=V$. \textcolor{black}{For example, for the attribute of gender, users only belong to one group, while users may belong to multiple groups in terms of the attribute of hobbies}. Given a set of groups $R$ and a model of information diffusion, the chief aim of the FIM problem is to determine a subset $S$ of $k$ nodes, denoted as seed set, to maximize the spread of influence while ensuring that different groups are affected fairly. The influence spread of $S$ is the total number of influenced nodes at the end of the information diffusion process. Tsang \emph{et al.}~\cite{tsang2019group} proposed two measures to capture group fairness in IM, namely $maximin\; fairness$ and $diversity\; constraints$.
	
	\textcolor{black}{$Maximin\; Fairness$ ($MF$) aims at maximizing the minimum influence received by any group, as proportional to the size of the group. In other words, the main goal of $MF$ is to improve the conditions for the $least\; well\-/off$ groups.} The maximin fairness of seed set $S$ is defined as:
	\begin{equation} \label{eq:eq-mf}\small
		MF(S) = \mathop{min}\limits_{i}\frac{I_{G,R_i}(S)}{|R_i|}
	\end{equation}
	\noindent where $I_{G,R_i}(S)$ represents the expected number of nodes in group $R_i$ activated by $S$ and $|R_i|$ represents the number of nodes contained in group $R_i$. In addition, $diversity\; constraints$ emphasize that the influence each group receives should be at least as much as each group may accomplish by itself, which is related to its size. We adopt $Diversity\; Constraint\; Violations$ ($DCV$)~\cite{tsang2019group} \textcolor{black}{to evaluate the fairness of seed sets obtained by different algorithms, which can be calculated as follows:
	\begin{equation} \label{eq:eq-dcv}\small
		DCV(S) = \frac{\sum_{i=1}^{q}max\{\frac{I_{G[R_i]}(k_i)-I_{G,R_i}(S)}{I_{G[R_i]}(k_i)}, 0\}}{q}
	\end{equation}}
	\noindent where $k_i=\lceil k|R_i|/n \rceil$ is the number of seeds that would be allocated to group $R_i$, $q$ is the number of groups, $G[R_i]$ represents a subgraph derived from $G$ by nodes in $R_i$, $I_{G[R_i]}(k_i)$ represents the maximum expected number of nodes in $G[R_i]$ that can be activated given a seed set of size $k_i$ and $I_{G,R_i}(S)$ represents the expected number of nodes in group $R_i$ that can be activated by $S$. What is more, Tsang \emph{et al.}~\cite{tsang2019group} proposed the $Price\; of\; Fairness$ ($PoF$) to measure the cost of ensuring a fair dissemination of information. In particular, $PoF^f$ is the $PoF$ under group-fairness metrics $f$, which can be calculated as follows:
	\begin{equation} \label{eq:eq-pof}\small
		PoF^f = \frac{I_G^{OPT}}{I_G^f}
	\end{equation}
	\noindent where $I_G^{OPT}$ represents the optimal influence spread and $I_G^f$ represents the best achievable influence spread under the metrics $f$ of group fairness. Apparently, $PoF\geq 1$ and smaller values are more desirable.
	
	
	\subsubsection{Influence Diffusion Model}
	\textcolor{black}{In this paper, we adopt the Independent Cascade (IC) model~\cite{goldenberg2001talk} to simulate the information dissemination process, since this model is widely adopted in previous FIM studies~\cite{tsang2019group,khajehnejad2021adversarial,khajehnejad2022crosswalk,stoica2019fairness}. } Each node in the IC model has two possible states, active or inactive. Inactive nodes have a chance to become active, but not vice versa. Moreover, active nodes attempt to activate their inactive neighbors. It is worth noting that an active node has only one chance to activate its neighbors when it is activated. The information dissemination process can be described as follows:
	
	\begin{enumerate}
		\item[1)] In the first place, the status of the nodes in the seed set $S$ is set to active. The information diffusion process starts from $S_0 = S$ at round $t=0$. $S_t$ denotes the set of nodes activated by $S_{t-1}$ at step $t$.
		
		\item[2)] In the second place, activated nodes in $S_{t-1}$ will attempt to activate their inactive neighbors with probability $p$ at step $t$. Additionally, newly activated nodes are added to $S_t$, which will spread the information in the next step.
		
		\item[3)] Eventually, the process terminates if no new nodes are activated, i.e., $S_t$ is empty. Let $\xi$ be the set of nodes activated by $S$, then $\xi = \bigcup_{i=0}^tS_i$.
	\end{enumerate}
	
	
	\subsubsection{Evaluation Function}
	In order to evaluate the fairness and influence spread of the seed set $S$, it is essential to design a reasonable evaluation function. A straightforward idea is to consider the two goals of fairness and influence spread simultaneously. Nevertheless, it hasn't escaped our notice that the fairness metrics $MF$ indirectly captures the influence spread of the seed set $S$, which focuses on improving the conditions for the $least\; well\-/off$ groups. In other words, the larger the value of $MF$, the larger the value of influence spread. In addition, existing definitions of fairness have different concerns. It means that only one definition of fairness may not be enough to meet complex real-world needs. Consequently, we make use of both $MF$ and $DCV$ to define an evaluation function denoted as $F$, which can not only better measure fairness but also consider influence spread. \textcolor{black}{The proposed evaluation function $F$ can be calculated as follows:
	\begin{equation} \label{eq:eq-MD} \small
		F(S) = \lambda \cdot MF(S)- (1-\lambda) \cdot DCV(S)
	\end{equation}
	\noindent where $\lambda\in [0,1]$ is the weight parameter for balancing $MF$  and $DCV$.} The construction of $F$ is based on the fact that we favor the seed set $S$ with higher $MF$ and lower $DCV$. The larger values of $F$ are more desirable.

	\subsection{Related work}
	Domingos and Richardson~\cite{domingos2001mining} were the first to introduce IM as an algorithmic problem and proposed a heuristic method to find the initial node set. Soon afterwards, Kempe \emph{et al.}~\cite{kempe2003maximizing} proved that the IM problem is NP-hard under both the Linear Threshold (LT) model and the IC model. Furthermore, they proposed a standard greedy algorithm to solve it, which can find a solution with an approximate guarantee of $(1-(1/\mathrm{e})-\epsilon)$, where $\mathrm{e}$ is the base of natural logarithm and $\epsilon$ is the sampling error. Over the years, massive IM algorithms have been proposed to find the seed set that maximizes the spread of information in social networks.
	
	Among the numerous IM approaches, the community-based method~\cite{wang2010community,shang2017cofim,kumar2022identifying} has shown great advantages in addressing IM in terms of both efficiency and effectiveness. For example, Wang \emph{et al.}~\cite{wang2010community} proposed a Community-based Greedy Algorithm (CGA) to find seed set in mobile social networks, which encompasses two components: community detection and node selection. To begin with, they designed a community detection algorithm considering information diffusion to obtain the community structure in social networks. Then, they employed a dynamic programming algorithm to find the seed set based on the obtained community structure, which significantly reduces the time cost. Shang \emph{et al.}~\cite{shang2017cofim} developed a Community-based Framework for Influence Maximization (CoFIM) on large-scale social networks, which can be divided into two phases: seeds expansion and intra-community propagation. Both stages of CoFIM are designed based on the community structure, which makes it achieve great results on large-scale social networks. Kumar \emph{et al.}~\cite{kumar2022identifying} designed a Community structure with Integrated Features Ranking algorithm (CIFR) for identifying influential nodes. CIFR considers not only local optimality but also global optimality when selecting influential nodes, which leads to high influence spread of the seed set.
	
	Apart from these community-based IM algorithms, there are numerous community-based approaches~\cite{li2018community,zhang2021local,styczen2022targeted} for tackling a great deal of variants of IM. For example, Styczen \emph{et al.}~\cite{styczen2022targeted} proposed Community and Gender-Aware Seeding (CG\footnotesize A\normalsize S) to solve the disparity influence maximization problem, which utilizes the information of community structure and its gender potential to select nodes with the purpose of iteratively modifying the seed set. Zhang \emph{et al.}~\cite{zhang2021local} designed a local-global Influence Indicator-based Constrained Evolutionary Algorithm (IICEA) for budgeted influence maximization, which takes into account both the local neighbor information and the community structure information. Li \emph{et al.}~\cite{li2018community} proposed an efficient Community-based Seeds Selection (CSS) algorithm for the location aware influence maximization problem. The CSS algorithm calculates users' community-based influence in advance, which is a key part of building PR-tree based indexes. CSS can efficiently find target users by traversing the PR-tree from the root in depth-first order.
	
	A new variant of influence maximization, called FIM, has recently received a lot of attention due to its practical significance~\cite{tsang2019group,stoica2019fairness,khajehnejad2021adversarial,khajehnejad2022crosswalk,razaghi2022group,gong2023influence,farnad2020unifying,rahmattalabi2021fair,becker2021fairness}. For example, Tsang \emph{et al.}~\cite{tsang2019group} first studied the FIM problem and proposed two notions of fairness: $maximin\; fairness$ and $diversity\; constraints$. In addition, they proved that the objective function of FIM is non-submodular under both proposed fairness concepts, implying that FIM cannot be easily solved using conventional greedy algorithms. To address this issue, they showed that the FIM problem can be reduced to the multi-objective submodular optimization problem under both proposed fairness concepts, and introduced an improved algorithm (DC) for general multi-objective submodular optimization. Stoica \emph{et al.}~\cite{stoica2019fairness} proposed two definitions of fairness, including $individual\; fairness$ and $group\; fairness$. Farnad \emph{et al.}~\cite{farnad2020unifying} proposed a flexible framework for FIM, which is based on the integer programming formulation of the FIM problem. Rahmattalabi \emph{et al.}~\cite{rahmattalabi2021fair} proposed a framework based on the social welfare theory for FIM, which can control the trade-off of fairness and efficiency through a parameter. Becker \emph{et al.}~\cite{becker2021fairness} recommended using randomization as a means to achieve fairness. Khajehnejad \emph{et al.}~\cite{khajehnejad2021adversarial} introduced an embedding-based approach (FairEmb) to tackle the FIM problem. Specifically, they jointly trained an auto-encoder for graph embedding and a discriminator for identifying sensitive attributes. However, this approach only works for networks with two groups, while users can be divided into more than two groups in most cases. Khajehnejad \emph{et al.}~\cite{khajehnejad2022crosswalk} proposed an embedding-based method (CrossWalk) to enhance the fairness of the node embedding, which is available for various random-walk based network embedding algorithms. \textcolor{black}{Due to the conflict between fairness and influence spread~\cite{tsang2019group}, multi-objective optimization-based methods~\cite{ma2022adaptive} become a good choice to solve this problem. For example, Gong \emph{et al.}~\cite{gong2023influence} proposed a multi-objective genetic approach called FIMMOGA to tackle the FIM problem. Razaghi \emph{et al.}~\cite{razaghi2022group} designed a multi-objective metaheuristic called SetMOGWO to address the FIM problem, which is developed based on the multi-objective gray wolf optimizer.}
	
	
	While many algorithms have been proposed to solve the FIM problem, these algorithms still suffer from striking a good balance between fairness and running time. For example, DC~\cite{tsang2019group} can find nodes that spread information fairly among different groups, but it owns high computational overhead. On the contrary, FairEmb~\cite{khajehnejad2021adversarial} and CrossWalk~\cite{khajehnejad2022crosswalk} can find solutions quickly at the expense of accuracy. In addition, none of these algorithms exploit the community structure information of social networks to improve fairness and reduce computational overhead. To address these issues, we propose a community-based evolutionary algorithm called CEA-FIM for FIM, which makes full use of community structure information to achieve a good balance between effectiveness and efficiency.

	\section{The Proposed Algorithm CEA-FIM}\label{sec:algo}
\begin{figure}[!htb]\vspace{-0.0cm}
		\begin{center}
			\centering
			\subfigure{
				\includegraphics[scale=0.31]{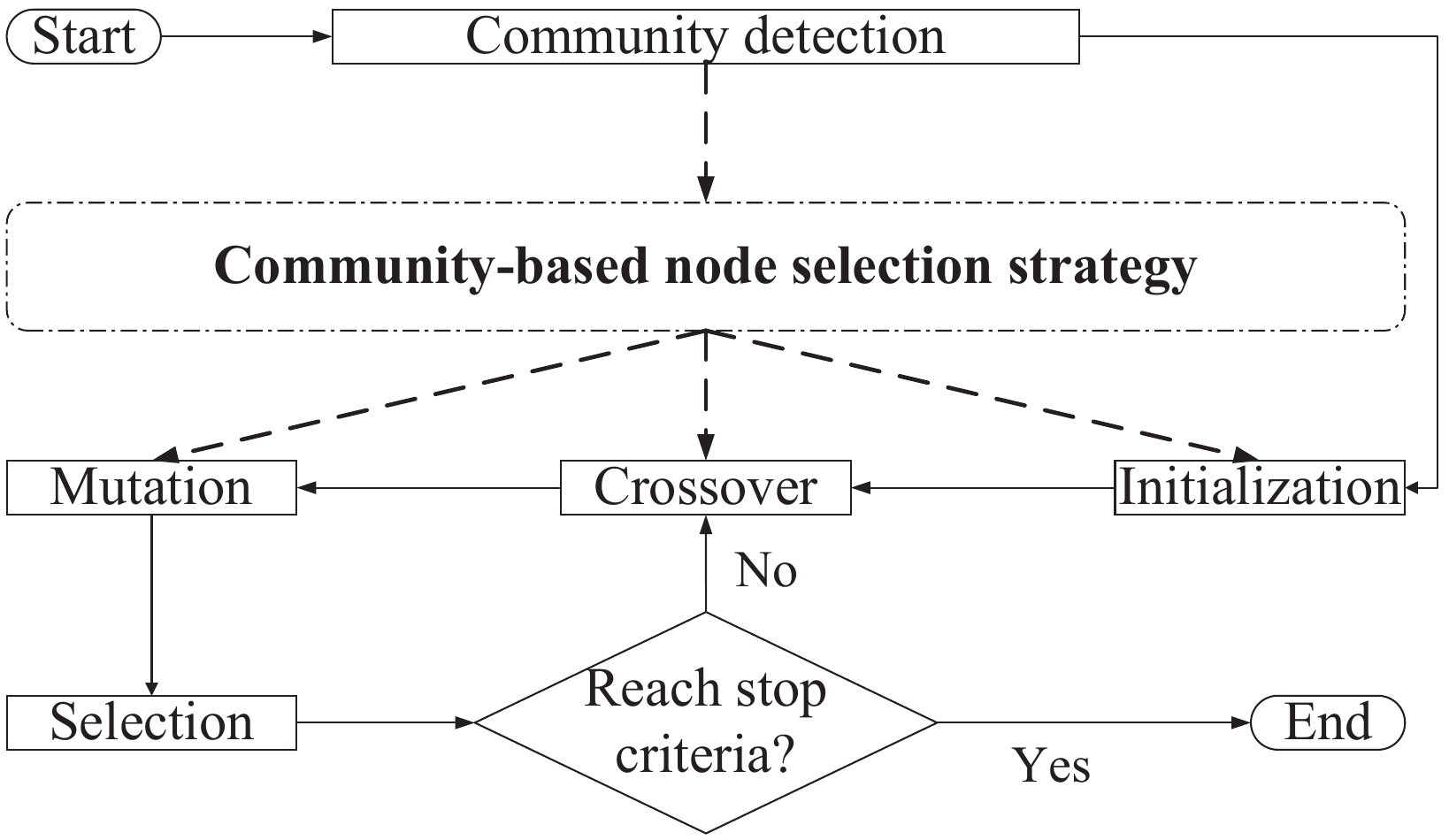}
			}
			\caption{The whole framework of the proposed algorithm CEA-FIM.}\vspace{-0.4cm}
			\label{fig:framework}
		\end{center}
	\end{figure}

	In this section, we propose a community-based evolutionary algorithm for the FIM problem, as is illustrated in Fig.~\ref{fig:framework}. Firstly, we design a fairness-aware node selection strategy based on the community structure. We, then, introduce an evolutionary algorithm CEA-FIM based on the proposed community-based node selection strategy, including the suggested initialization, crossover and mutation strategies. \textcolor{black}{The computational complexity of CEA-FIM can be found in Section A of Supplementary Materials}. Details descriptions of CEA-FIM are provided in the following subsections. Table~\ref{tab:notations} enumerates key notations used in this paper.
	
	\begin{table}[ht]
		\centering
		\begin{scriptsize}
			\caption{Notations used in this paper.}\label{tab:notations} \vspace{-0.2cm}
			\begin{tabular}{l l}
				\hline
				Notation &Definition\\
				\hline
				$G=(V,E)$ &Social network $G$ with node set $V$ and edge set $E$.\\
				$C_i$ &The $i$th community\\
				$v_i$ &The $i$th node in $V$\\
				$k$ &The size of seed set $S$\\
				$m$ &The number of communities\\
				$n$ &The number of nodes in $G$\\
				$w$ &\textcolor{black}{The number of edges in $G$}\\
				$q$ &The number of groups\\
				$pop$ &The size of population\\
				$g_{max}$ &The maximum number of iterations\\
				$cr$ &The probability of crossover\\
				$mu$ &The probability of mutation\\
				$P_i$ &The $i$th individual of population $P$\\
				$SC_i$ &The community score of community $C_i$\\
				$SN_i$ &The node score of node $v_i$\\
				$A_i$ &The number of nodes with the $i$th attribute in $G$\\
				$u_i$ &The urgency of the $i$th attribute\\
				$AC_i$ &The node attributes contained in community $C_i$\\
				$CA_i$ &\makecell[l]{The number of nodes with the $i$th attribute in the communities \\to which the selected nodes belong}\\
				\hline
			\end{tabular}
		\end{scriptsize}
	\end{table}\vspace{-0.1cm}
	
	\subsection{Community-based Node Selection}
	
	\begin{figure*}[htb]\vspace{-0.7cm}
		\small
		\begin{center}
			\centering
			\subfigure{
				\includegraphics[width=7.0in]{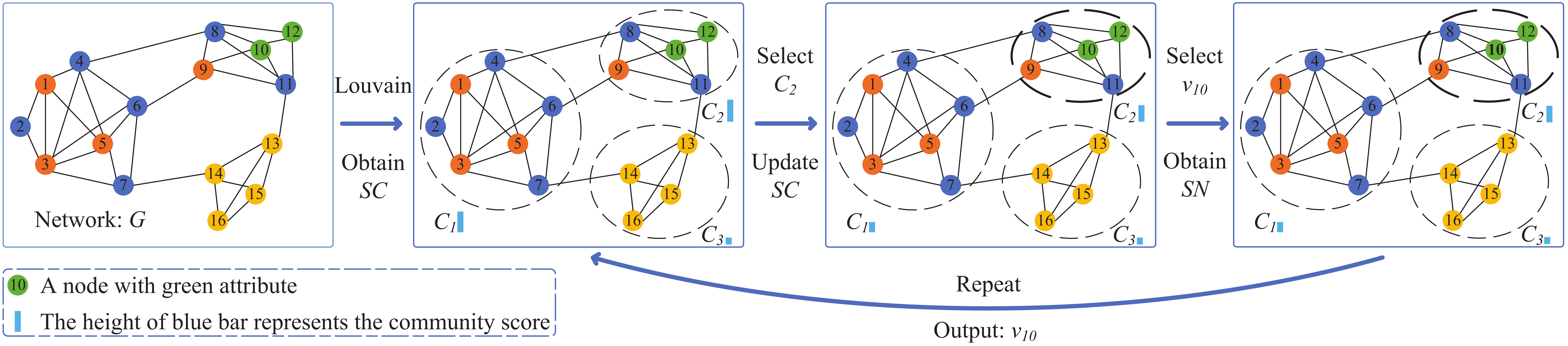}
			} \vspace{-0.2cm}
			\caption{A schematic illustration of the community-based node selection process, \textcolor{black}{where node $v_{10}$ is selected from community $C_2$}.}\label{fig:cs-example}\vspace{-0.3cm}
		\end{center}
	\end{figure*}\vspace{-0.0cm}
	
	In this section, we introduce an effective node selection strategy, which exploits the community structure of social network. The reason why we consider community structure information can be attributed to two points. On the one hand, FIM seeks to find a seed set for fair information dissemination without losing maximimum influence spread. However, traditional node selection strategies, such as degree-based node selection strategy, are difficult to effectively obtain useful nodes in a short time. On the other hand, community structure is an essential characteristic of social networks. Nodes can be divided into several communities, with dense connections within communities and sparse connections between communities. Apparently, nodes in a community can influence each other easily. Ideally, one node can even influence all nodes in the community. As a result, the excessive cost of finding potential nodes over the whole network would be reduced tremendously if we use the information of community structure. \textcolor{black}{In our proposed node selection strategy, every node in the network has a chance to be selected, while the chances are not equal. More specifically, potential nodes get more attention than other nodes due to the utilization of community structure information.}
	
	To effectively find the potential nodes, we propose a community-based node selection strategy. The key idea is to first select a potential community and then select a potential node from it. Specifically, we first obtain the community structure $C=\{C_1,C_2,...,C_m\}$ using a community detection algorithm. Afterwards, each community $C_i$ in $C$ is assigned a score $SC_i$, where $SC_i$ is defined as:
	\begin{equation} \label{eq:eq-cs} \small
		SC_i = |C_i|\cdot\sum_{j\in{AC_i}}u_j
	\end{equation}
	\noindent where $|C_i|$ represents the number of nodes in community $C_i$, $AC_i$ represents the node attributes contained in community $C_i$ and $u_j$ represents the urgency of the $j$th attribute. More specifically, $u_j$ is the $attribute\; urgency$ defined based on the selected nodes and community structure information. The urgency of the $j$th attribute is calculated by Eq.~\eqref{eq:eq-urgency}.
	\begin{equation} \label{eq:eq-urgency} \small
		u_j = \mathrm{e}^{-\frac{CA_j}{A_j}}
	\end{equation}
	\noindent where $\mathrm{e}$ is the base of natural logarithm, $CA_j$ represents the number of nodes with the $j$th attribute in the communities to which the selected nodes belong and $A_j$ represents the number of nodes with the $j$th attribute in network $G$. $CA_j$ is defined as:
	\begin{equation} \label{eq:eq-ca} \small
		CA_j = \sum_{t=1}^{m}C_{tj}
	\end{equation}
	\noindent where $C_{tj}$ represents the number of nodes with the $j$th attribute included in community $C_t$. Note that if no node in community $C_t$ is selected or there is no node with the $j$th attribute in community $C_t$, the value of $C_{tj}$ is 0. It is clearly that the community score $SC$ takes into consideration both the size of community and the node attributes contained in community. Only in this way can we identify potential communities, where nodes can spread information fairly without losing too much influence spread. In addition, $attribute\; urgency$ is used to measure the degree to which nodes of various attributes are influenced. The greater the $attribute\; urgency$ value of a certain attribute, the greater the bias that the node with this attribute may receive. Owing to the fact that the nodes in the community are densely connected, we roughly assume that any node can influence other nodes in the community. Once a node is selected, the $attribute\; urgency$ of the node attributes contained in its community will be reduced according to the number of nodes with different attributes. Most notably, the community score varies with the $attribute\; urgency$. After calculating the community score, the probability of community $C_i$ being selected can be calculated as follows:
	\begin{equation} \label{eq:Pci} \small
		p(C_i) = \frac{SC_i}{\sum\limits_{t=1}^{m}SC_t}
	\end{equation}
	\noindent where $SC_i$ represents the community score of community $C_i$. According to the probability of the community being selected, we can find a potential community without difficulty. Significantly, the score of community will be updated, if the potential community is selected for the first time.
	
	Subsequently, in the selected community, we assume that the nodes with diverse attributes are uniformly distributed due to lack of prior knowledge. As a consequence, we should find the node in the center of the community, which can greatly influence the whole community. Nevertheless, evaluating nodes only within the community may achieve poor results, especially in a network with a high average degree where there are many connections between communities. To address this issue, we employ the PageRank~\cite{page1999pagerank} algorithm to calculate the score of the node, denoted as $SN$, over the whole network. Let $SN_{i,\tau}$ denote the score of node $v_i$ at iteration $\tau$ and it will be updated as follows:
	\begin{equation} \label{eq:ns} \small
		SN_{i,\tau+1} = d \cdot \sum\limits_{v_j \in H_{i}^{in}}\frac{SN_{j,\tau}}{|H_{j}^{out}|}+\frac{1-d}{n}, \forall v_i\in V
	\end{equation}
	\noindent where \textcolor{black}{$n$ is the number of nodes in $G$}, $d$ is the damping coefficient to prevent the node score from increasing indefinitely, $H_{i}^{in}$ is the set of in-degree neighbors of node $v_i$, $|H_{i}^{out}|$ is the number of out-degree neighbors of node $v_i$ and $n$ is the number of nodes in $G$. For nodes in potential community $C_j$, the probability of node $v_i$ being selected can be calculated as follows:
	\begin{equation} \label{eq:Pni} \small
		p(v_i) = \frac{SN_i}{\sum\limits_{t=1}^{|C_j|}SN_t}
	\end{equation}
	\noindent where $SN_i$ represents the node score of node $v_i$ and $|C_j|$ represents the number of nodes in $C_j$. Based on the corresponding probability of these nodes, a potential node can be easily selected.
	
	The procedure of the proposed community-based node selection strategy is illustrated in Fig.~\ref{fig:cs-example}. In this study, we adopt a simple yet effective Louvain algorithm~\cite{blondel2008fast} to obtain the community structure $C$ of network $G$. Based on the community structure, the score of each community can be calculated using Eq.~\eqref{eq:eq-cs}, which is shown as blue bar of varying heights in Fig.~\ref{fig:cs-example}. There are 7 nodes in community $C_1$, and the number of node attribute is 2. While community $C_2$ barely contains 5 nodes, the number of node attribute is 3. After calculation, the community score of $C_1$ is 14 and that of $C_2$ is 15. It indicates that $C_2$ deserves more attention than $C_1$ in terms of fair influence spread. Assume $C_2$ is selected according to Eq.~\eqref{eq:Pci}, the community scores of $C_1$ and $C_2$ should be updated, while that of $C_3$ remains unchanged. Last but not least, node $v_{10}$ is selected according to Eq.~\eqref{eq:Pni}, which will spread information fairly across the network $G$.

	\subsection{Optimal Solution Search}
	
	\subsubsection{Overall Framework}
	
	\begin{algorithm}[t] \label{algo:framework}
		\begin{small}
			\begin{algorithmic}[1]
				\caption{The framework of CEA-FIM \label{algo:main}}
				\REQUIRE{$G = (V,E)$: social network; $pop$: the size of population; $g_{max}$: the maximum number of iterations; $cr$: the probability of crossover; $mu$: the probability of mutation; $k$: the size of the seed set;}
				\ENSURE{A $k$-node set $S$ with fair influence spread;}
				
				\textbf{Step1: Community Detection.}
				\STATE $C=\{C_1,C_2,...,C_{m}\} \leftarrow$ obtain community structure by the Louvain algorithm;
				
				\textbf{Step2: Population Initialization.}
				\STATE $SN=\{SN_1,SN_2,...,SN_{n}\} \leftarrow$ calculate the scores of nodes in $G$ using Eq.~\eqref{eq:ns};
				\STATE $P \leftarrow$ $\underline{Initialization(G,pop,k,C,SN)}$;
				
				\textbf{Step3: Population Evolution.}
				\STATE Set iteration counter to zero: $g$ = 0;
				\WHILE {$g < g_{max}$}
					\STATE Sort all individuals of $P$ in descending order using $F$;
					\STATE $P'\leftarrow$ $\underline{Crossover(G,pop,k,C,SN,P,cr)}$;
					\STATE $P''\leftarrow$ $\underline{Mutation(G,pop,k,C,SN,P',mu)}$;
					\FOR{$i = 1$ to $pop$}
						\STATE $P_i\leftarrow$ $Max(P_i, P''_i, F)$;
					\ENDFOR
					\STATE $g = g+1$;
				\ENDWHILE
				\STATE $S\leftarrow$ $\arg max_{P_i \in P}(F(P_i))$;
			\end{algorithmic}
		\end{small}
	\end{algorithm}
	
	In this paper, we adopt a $k$-bit real variable $P_i$ to represent each individual of the population $P$. More precisely, we let $P_i=\{P_{ij}\in [1,2,...,n]\}_k,\forall j\neq t,P_{ij}\neq P_{it}$, where $P_{ij}=x$ represents that node $v_x$ is the $j$th element of $P_i$. It is noteworthy that the elements in each individual are unique.
	
	Algorithm~\ref{algo:framework} shows the framework of the proposed method, which consists of three key steps: community detection, population initialization and population evolution. In the first step, the network community structure is detected by the Louvain algorithm~\cite{blondel2008fast}, which is an essential element for subsequent operations. In the second step, the scores of nodes in $G$ are calculated in preparation for population initialization and evolution. Thereafter, population $P$, which utilizes the information of the community structure, is initialized with diversity and convergence (see line 3). In the third step, all individuals of $P$ are sorted in descending order using proposed evaluation function $F$. For the sorted population $P$, the proposed crossover operator is performed to generate high-quality solutions, which takes advantage of superior solutions to accelerate the convergence of poor solutions. Furthermore, the proposed mutation operator is executed to find the optimal solution. Significantly, both the proposed crossover and mutation operators incorporate the suggested community-based node selection strategy. At each iteration, each individual $P_i$ in $P$ generates a new individual $P''_i$ through the proposed crossover and mutation strategy (see lines 7-8). In the end, the individual with a higher evaluation function value between $P_i$ and $P''_i$ will survive. In the following subsections, we will introduce these critical steps in detail.
	
	\subsubsection{Initialization}
	
	\begin{algorithm}[t] \label{algo:init}
		\begin{small}
			\begin{algorithmic}[1]
				\caption{Initialization($G$, $pop$, $k$, $C$, $SN$)}
				\REQUIRE{$G = (V,E)$: social network; $pop$: the size of population; $k$: the size of the seed set; $C$: community structure; $SN$: node scores;}
				\ENSURE{An initialized population $P$;}
				\STATE Initialize $P$ with empty vectors;
				\FOR {$i = 1$ to $pop$}
					\STATE $SC=\{SC_1,SC_2,...,SC_{m}\} \leftarrow$ calculate the scores of communities in $C$ using Eq.~\eqref{eq:eq-cs};
					\STATE Initialize community counter $CC=\{0,0,...,0\}_m$;
				\FOR {$j = 1$ to $k$}
					\STATE Obtain the potential community $C_t$ according to the probability calculated by Eq.~\eqref{eq:Pci};
					\STATE Update SC using Eq.~\eqref{eq:eq-cs};
					\STATE $CC_t = CC_t+1$;
				\ENDFOR
				\FOR {$h = 1$ to $m$}
					\IF {$CC_h \neq 0$}
						\STATE Add the top $CC_h$ nodes with the highest node scores in $C_h$ to $P_i$;
					\ENDIF
				\ENDFOR
				\ENDFOR
				
			\end{algorithmic}
		\end{small}
	\end{algorithm}
	
	\begin{figure}[tb]\vspace{-0.2cm}
		\begin{center}
			\centering
			\subfigure{
				\includegraphics[scale=0.3]{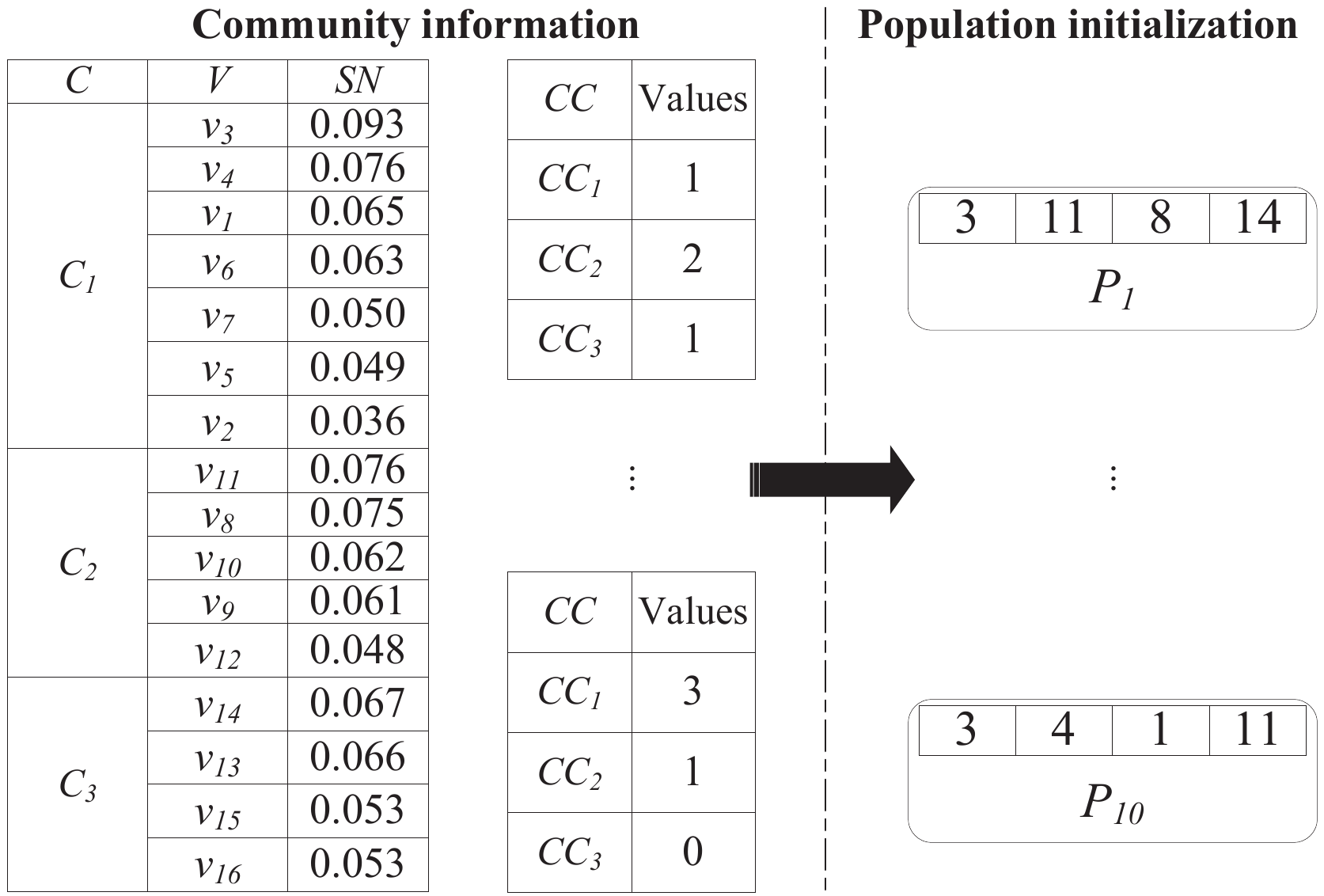}
			}
			\caption{Illustration of the community-based population initialization process.}\vspace{-0.3cm}
			\label{fig:initialization}
		\end{center}
	\end{figure}
	
	In order to initialize high-quality solutions, we design a community-based initialization strategy as shown in Algorithm~\ref{algo:init}, \textcolor{black}{in which each node has a chance to be selected.} In line 1, we initialize a population $P$ consisting of $pop$ individuals of size 0. In line 3, the community score $SC$ is calculated using Eq.~\eqref{eq:eq-cs}. In line 4, we initialize a community counter $CC$, whose $h$th element represents the number of times the $h$th community has been selected. In lines 5-9, we repeatedly  select $k$ communities according to Eq.~\eqref{eq:Pci}. It is noteworthy that if the potential community $C_t$ is selected for the first time, the community score $SC$ needs to be updated. The number of times each community is selected is recorded in $CC$. Eventually, in lines 10-14, the top $CC_h (h\in \{1,2,...,m\})$ nodes with the highest scores in each community are added to the individual $P_i$. In this way, the obtained initial population includes a host of high-quality solutions, which provides a good foundation for further optimal solution search.
	
	An intuitive example based on the network $G$ \textcolor{black}{(shown in Fig.~\ref{fig:cs-example})} can be found in Fig.~\ref{fig:initialization}. Four nodes are selected from social network $G$ \textcolor{black}{to form an individual.} Detailed community information of $G$ can be found on the left panel, and the initial population $P$ is shown on the right panel. The community counter $CC$ records the number of times each community is selected. In community $C_i$, we select the top $CC_i$ nodes with the highest scores, i.e., $SN$, to initialize the individual. Taking the community counter corresponding to individual $P_1$ as an example, the times of $C_1$, $C_2$ and $C_3$ being selected are 1, 2 and 1 respectively. Thus, nodes $v_3$, $v_{11}$, $v_8$ and $v_{14}$ are selected to form individual $P_1$.
	
	\subsubsection{Crossover}
	
	\begin{algorithm}[t] \label{algo:crossover} \vspace{-0.0cm}
		\begin{small}
			\begin{algorithmic}[1]
				\caption{Crossover($G$, $pop$, $k$, $C$, $SN$, $P$, $cr$)}
				\REQUIRE{$G = (V,E)$: social network; $pop$: the size of population; $k$: the size of the seed set; $C$: community structure; $SN$: node scores; $P$: population; $cr$: the probability of crossover;}
				\ENSURE{A new population $P'$;}
				\FOR{$i = 1$ to $pop/2$}
					\FOR{$j = 1$ to $k$}
						\IF {$rand(1) < cr$}
							\STATE Swap $P_{ij}$ and $P_{(pop-i+1)j}$ with each other;
						\ENDIF
					\ENDFOR
				\ENDFOR
				\FOR{$i = 1$ to $pop$}
					\STATE Remove duplicate elements in $P_i$;
					\IF {$|P_i|!=k$}
						\STATE $SC=\{SC_1,SC_2,...,SC_{m}\} \leftarrow$ calculate the scores of communities in $C$ using Eq.~\eqref{eq:eq-cs};
						\WHILE {$|P_i|<k$}
							\STATE Obtain the potential community $C_t$ according to the probability calculated by Eq.~\eqref{eq:Pci};
							\STATE Update SC using Eq.~\eqref{eq:eq-cs};
							\STATE Obtain the potential node $v_s$ from $C_t$ according to the probability calculated by Eq.~\eqref{eq:Pni};
							\STATE Add node $v_s$ to $P_i$;
						\ENDWHILE
					\ENDIF
				\ENDFOR
			\end{algorithmic}
		\end{small}
	\end{algorithm}

	After obtaining the initial population, as described in Algorithm~\ref{algo:crossover}, the proposed crossover operator is adopted to search for better solutions. Specifically, in lines 1-7, we perform a uniform crossover between the $i$th individual and the $(pop-i+1)$th individual. Note that duplicate elements in each individual should be removed. In lines 8-19, we ensure that there are $k$ unique elements in each individual. To be specific, we first remove duplicate elements in each individual. Then, we utilize the proposed community-based node selection strategy to add nodes to individuals. In the end, many excellent individuals will be produced after the proposed crossover operation, which prevents the population from falling into local optimum.
	
\begin{figure}[!tb]\vspace{-0.2cm}
		\begin{center}
			\centering
			\subfigure{
				\includegraphics[scale=0.36]{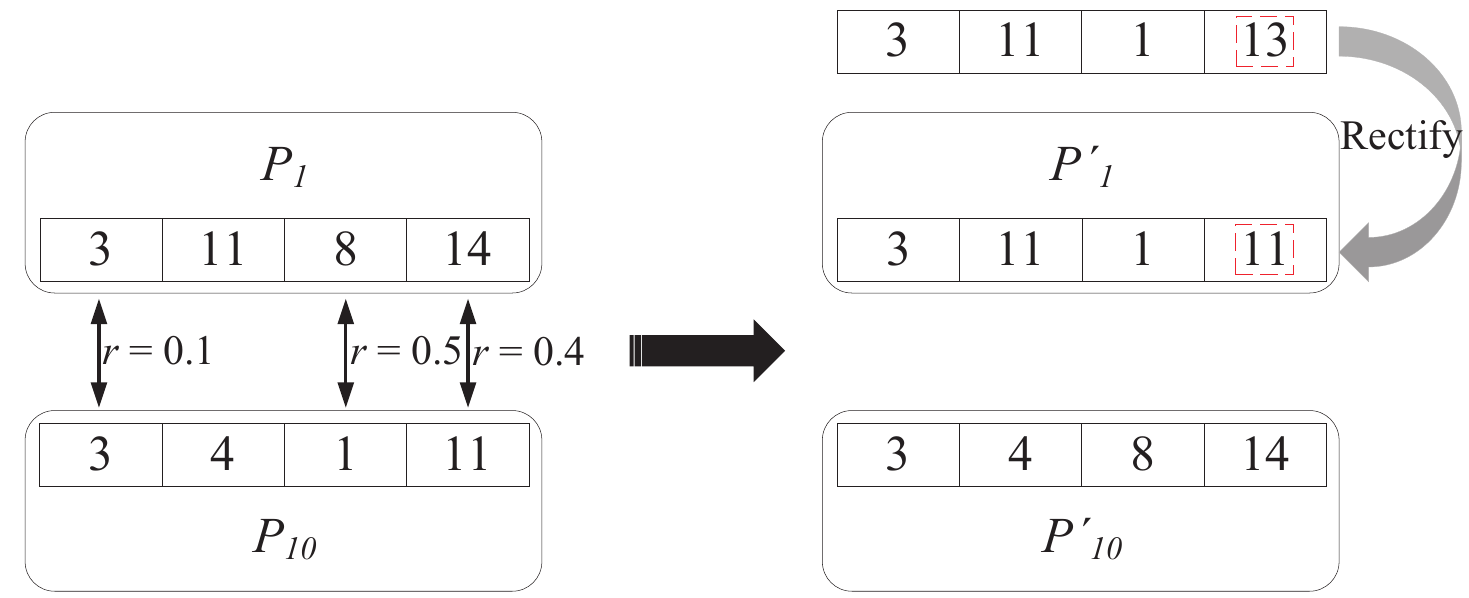}
			}
			\caption{Illustration of the community-based uniform crossover operator when the crossover probability $cr=0.6$.}\vspace{-0.3cm}
			\label{fig:crossover}
		\end{center}
	\end{figure}

	Fig.~\ref{fig:crossover} shows a vivid example based on the population $P$ in Fig.~\ref{fig:initialization}. $P_1$ is the best individual in the population, while $P_{10}$ is the worst. $P_1$ and $P_{10}$ randomly exchange elements at the same position if the random number $r$ is smaller than the crossover probability $cr$ and generate two new individuals $P'_1$ and $P'_{10}$. Note that duplicate node $v_{11}$ exits in $P'_1$, so we perform the proposed community-based node selection strategy to select a unique node $v_{13}$ to rectify $P'_1$.

	\subsubsection{Mutation}
	
	\begin{algorithm}[t] \label{algo:mutation}
		\begin{small}
			\begin{algorithmic}[1]
				\caption{Mutation($G$, $pop$, $k$, $C$, $SN$, $P'$, $mu$)}
				\REQUIRE{$G = (V,E)$: social network; $pop$: the size of population; $k$: the size of the seed set; $C$: community structure; $SN$: node scores; $P'$: population; $mu$: the probability of mutation;}
				\ENSURE{A new population $P''$;}
				\FOR{$i = 1$ to $pop$}
					\FOR{$j = 1$ to $k$}
						\IF {$rand(1) < mu$}
							\STATE $SC=\{SC_1,SC_2,...,SC_{m}\} \leftarrow$ calculate the scores of communities in $C$ using Eq.~\eqref{eq:eq-cs};
							\STATE Obtain the potential community $C_t$ according to the probability calculated by Eq.~\eqref{eq:Pci};
							\STATE Obtain the potential node $v_s$ from $C_t$ according to the probability calculated by Eq.~\eqref{eq:Pni};
							\STATE Replace element $P'_{ij}$ with node $v_s$;
						\ENDIF
					\ENDFOR
				\ENDFOR
			\end{algorithmic}
		\end{small}
	\end{algorithm}

	\begin{figure}[!htb]\vspace{-0.2cm}
		\begin{center}
			\centering
			\subfigure{
				\includegraphics[scale=0.38]{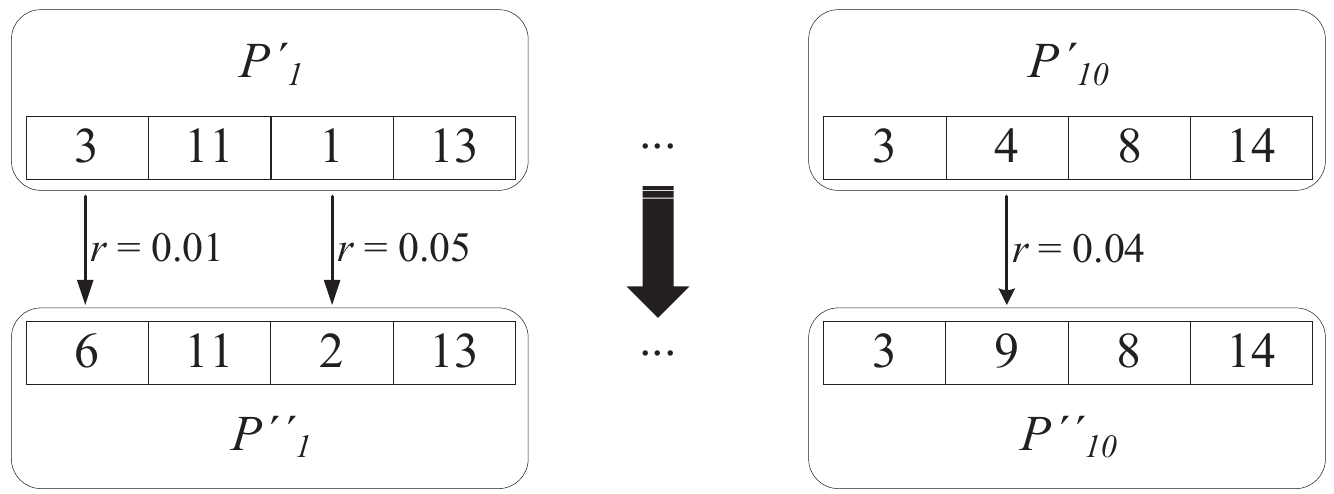}
			}
			\caption{Illustration of the community-based random mutation operator when the mutation probability $mu=0.1$.}
			\label{fig:mutation}
		\end{center}
	\end{figure}
	
	To explore better solutions, we design a random mutation strategy based on community information as shown in Algorithm~\ref{algo:mutation}. Each element of the individual will mutate with probability $mu$. More specifically, when we mutate the $j$th element of individual $P'_i$, we calculate the community score in line 4, which does not take the $j$th element into account. We then select the community $C_t$ and the node $v_s$ in turn, and replace $P'_{ij}$ with the newly selected node $v_s$. \textcolor{black}{Node $v_s$ is a high potential node selected by the proposed community-based node selection strategy}. Note that duplicate nodes are not allowed in individual $P'_i$. Finally, a new population $P''$ containing many elite individuals is generated.
	
	Fig.~\ref{fig:mutation} gives an illustration of the proposed mutation operator. Each element in the individual will mutate if the random number $r$ is less than $mu$. In the example, only node $v_4$ in individual $P'_{10}$ is mutated, because its corresponding random number $r$ is 0.04, which is smaller than the mutation probability $mu$. What's more, the newly added node $v_9$ is generated based on proposed community-based node selection strategy.

	\section{Expriments}\label{sec:expe}
	This section verifies the effectiveness and efficiency of the proposed algorithm CEA-FIM by comparing it with state-of-the-art baseline algorithms and its variant on real-world and synthetic networks. Specifically, the experimental setting is introduced in Section~\ref{subsec:setting}, and the experimental results and analysis are presented in Section~\ref{subsec:result}.
	
	\subsection{Experimental Setting}\label{subsec:setting}
	
	\subsubsection{Datasets}

	\begin{table}[ht]\vspace{-0.5cm}
		\centering
		\begin{scriptsize}
			\caption{The real-world social networks with different characteristics.}\label{tab:real-datasets} \vspace{-0.0cm}
			\begin{tabular}{|c|c|c|c|c|}
				\hline
				Networks  & \#Nodes   &\#Edges  &Attributes  &\#Attributes \\
				\hline
				Rice-Facebook & 1,205& 42,443&  age& 2\\
				\hline
				Twitter & 3,560& 6,677&  political learning& 3\\
				\hline
			\end{tabular}
		\end{scriptsize}
	\end{table}\vspace{-0.0cm}

	\begin{table}[ht]\vspace{-0.2cm}
		\centering
		\setlength{\tabcolsep}{3.27mm}{
		\begin{scriptsize}
			\caption{The synthetic networks with different characteristics.}\label{tab:syn-datasets} \vspace{-0.0cm}
			\begin{tabular}{|c|c|c|c|c|}
				\hline
				Networks  & \#Nodes   &\#Edges  &Attributes  &\#Attributes \\
				\hline
				Synth1& 500& 1,689&\makecell[c]{region\\ethnicity\\age\\gender} & \makecell[c]{13\\5\\7\\2}\\
				\hline
				Synth2& 500& 3,668&  -& 2\\
				\hline
				Synth3& 500& 2,918&  -& 3\\
				\hline
			\end{tabular}
		\end{scriptsize}
		}
	\end{table}\vspace{-0.0cm}
	
	In our experiments, we adopt  two real-world social networks and three synthetic networks with different characteristics. All datasets are undirected graphs and collected from related work~\cite{tsang2019group,khajehnejad2022crosswalk}, \textcolor{black}{where real-world social networks are subgraphs.} The statistical information about these datasets are shown in Table~\ref{tab:real-datasets} and Table~\ref{tab:syn-datasets}, where \#Nodes and \#Edges respectively represent the number of nodes and edges in the network, Attributes and \#Attributes denote attributes and number of attributes of nodes in the network respectively. Details about these networks are introduced as follows:
	
	\begin{enumerate}
		\item[1)] Rice-Facebook dataset: This real-world dataset is a sub-graph of the Rice University Facebook dataset~\cite{mislove2010you}, which represents the friendship relations between students of Rice University. It contains 1,205 nodes and 42,443 edges. The information of students is integrated into nodes in this graph, including college, age, and major. We consider the age of students as the sensitive attribute. Students whose age is greater than 20 are excluded. Then, we divide the nodes whose age attribute is 20 into group A, and the rest into group B. Group A has 344 nodes, while group B has 97 nodes. The number of intra-group and inter-group connections are $e_{intra}^A=7,441$, $e_{intra}^B=513$ and $e_{inter}^{AB}=1,779$, \textcolor{black}{where $e_{intra}^A$ represents the number of connected edges within group A, and $e_{inter}^{AB}$ represents the number of connected edges between groups A and B.}
		
		\item[2)] Twitter dataset: This real-world dataset is a sub-graph of the Twitter dataset~\cite{babaei2016efficiency,cha2010measuring} with 3,560 nodes. All nodes are divided into three groups according to their political learning: group A (neutrals) with 2,598 nodes, group B (liberals) with 782 nodes and group C (conservatives) with 180 nodes. The number of intra-group and inter-group connections are $e_{intra}^A=3,724$, $e_{intra}^B=950$, $e_{intra}^C=74$, $e_{inter}^{AB}=1,461$, $e_{inter}^{AC}=359$ and $e_{inter}^{BC}=109$.
		
		\item[3)] Synth1 dataset: This synthetic dataset is used to model an obesity prevention intervention in the Antelope Valley region of California~\cite{wilder2018optimizing} with 500 nodes. Each node in the network has four attributes, including geographic region, ethnicity, age and gender. According to each attribute, the nodes can be divided into thirteen, five, seven and two groups, respectively. We regard this dataset as four different datasets, denoted as Synth1-region, Synth1-ethnicity, Synth1-age and Synth1-gender.
		
		\item[4)] Synth2 dataset: This synthetic dataset~\cite{khajehnejad2022crosswalk} comprises two groups. There are 350 nodes in group A, and 150 nodes in group B. The intra-group and inter-group connection probabilities are $CP_{intra}^A=CP_{intra}^B=0.025$ and $CP_{inter}^{AB}=0.001$, \textcolor{black}{where $CP_{intra}^A$ represents the probability of forming an edge within group A, and $CP_{inter}^{AB}$ represents the probability of forming an edge between groups A and B.}
		
		\item[5)] Synth3 dataset: This synthetic dataset~\cite{khajehnejad2022crosswalk} comprises three groups. There are 300 nodes in group A, 125 nodes in group B and 75 nodes in group C. The intra-group connection probabilities are $CP_{intra}^A=CP_{intra}^B=CP_{intra}^C=0.025$, and the inter-group connection probabilities are $CP_{inter}^{AB}=0.001$ and $CP_{inter}^{AC}=CP_{inter}^{BC}=0.0005$.
		
	\end{enumerate}
	
	\subsubsection{Baseline Algorithms}
	In this paper, we compare CEA-FIM with \textcolor{black}{five} representative and \textcolor{black}{open-source-code} baseline algorithms and a variant of CEA-FIM, which are introduced as follows:
	\begin{enumerate}
		\item[1)] Greedy: It is a standard greedy algorithm for IM~\cite{kempe2003maximizing}, which maximizes the influence spread without considering fairness.
		
		\item[2)] DC: It is an improved algorithm for general multi-objective submodular maximization problems~\cite{tsang2019group}, which considers the diversity constraints.
		
		\item[3)] FairEmb: It is an adversarial graph embeddings approach~\cite{khajehnejad2021adversarial}, which co-trains an auto-encoder and a discriminator to make the embedding evenly distributed across sensitive attributes. This approach is only applicable to networks of 2 groups.
		
		\item[4)] CrossWalk: It is a fairness-enhanced node representation learning method~\cite{khajehnejad2022crosswalk}, which reweights special edges to enhance the fairness of random walk based node embeddings.
		
		\item[5)] \textcolor{black}{SetSOGWO: It is a single-objective version of SetMOGWO~\cite{razaghi2022group}, which is developed based on gray wolf optimization~\cite{roayaei2021binarization}. For a fair comparison, we use the same evaluation function (i.e., $F$ shown in Eq.~\eqref{eq:eq-MD}) as CEA-FIM to evaluate the seed set.}
		
		\item[6)] REA-FIM: It is a variant based on CEA-FIM, which replaces the proposed community-based node selection strategy with the random node selection strategy. Specifically, node selection in REA-FIM is random without utilizing any useful information. REA-FIM is used to verify the effectiveness of the proposed community-based node selection strategy.
		
	\end{enumerate}
	
	\subsubsection{Experimental Settings and Conditions}
	\begin{table}[ht]
		\centering
		\begin{scriptsize}
			\caption{\textcolor{black}{The parameter setting of CEA-FIM.}}\label{tab:parameter} \vspace{-0.2cm}
			\setlength{\tabcolsep}{2.8mm}{
				\begin{tabular}{|c|c|c|c|c|c|c|c|}
					\hline
					Parameters  &$k$ &$pop$ &$g_{max}$ &$cr$ &$mu$ &$\lambda$ &$p$ \\
					\hline
					Values & 40 & 10 &150 &0.6 &0.1 &0.5 &0.01\\
					\hline
			\end{tabular}}
		\end{scriptsize}
	\end{table}\vspace{-0.0cm}
	
	Just like all baseline algorithms, we utilize the IC model~\cite{goldenberg2001talk} as diffusion model. We set the propagation probability to be $p=0.01$ and fix $k=40$ seeds, which are also adopted by the algorithms FairEmb and CrossWalk. Following the setting of algorithm DC, we utilize a constant number (i.e., 1,000) of live-edge graphs to simulate the influence spread. For all baseline algorithms, we adopt the parameters suggested in their original papers. \textcolor{black}{As shown in Table~\ref{tab:parameter}, the parameters for the proposed CEA-FIM algorithm are set as follows: the number of generations $g_{max}=150$, the size of population $pop=10$, the crossover probability $cr=0.6$, and the mutation probability $mu=0.1$.
In addition, the weight parameter $\lambda$ for balancing $MF$  and $DCV$ is set to 0.5, and the experimental analysis of $\lambda$ can be found in Section B of Supplementary Materials.
Note that the population size and number of generations in SetSOGWO are set to the same values as in CEA-FIM for a fair comparison.} We report the averaged results of 10 runs of the algorithms for each network. In addition, the proposed algorithm SSR-PEA and other baseline methods are written in Python3.8 environment and run on a PC with Intel Core i7-10750H CPU @2.60GHz and 16.0 GB memory.

	\subsection{Experimental Results and Analysis}\label{subsec:result}
	
	\subsubsection{Comparison Results between CEA-FIM and Baselines}
	\begin{figure}[!htb] \vspace{-0.5cm}
		\begin{center}
			\centering
			\subfigure[]{
				\includegraphics[width=1.58in]{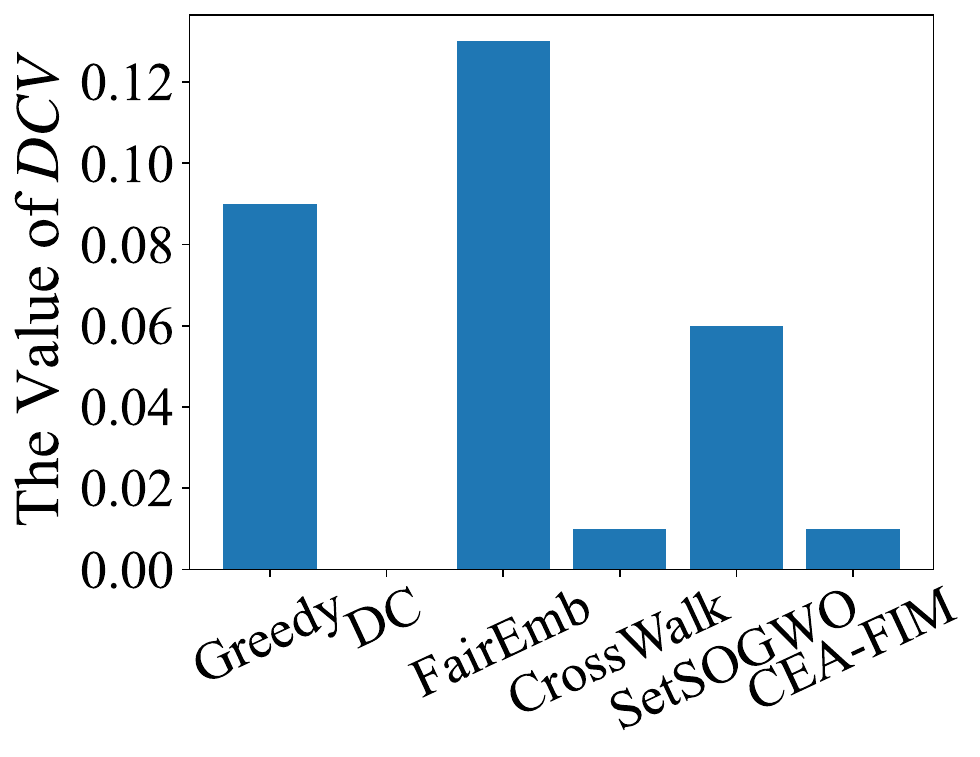}
			}
			\subfigure[]{
				\includegraphics[width=1.58in]{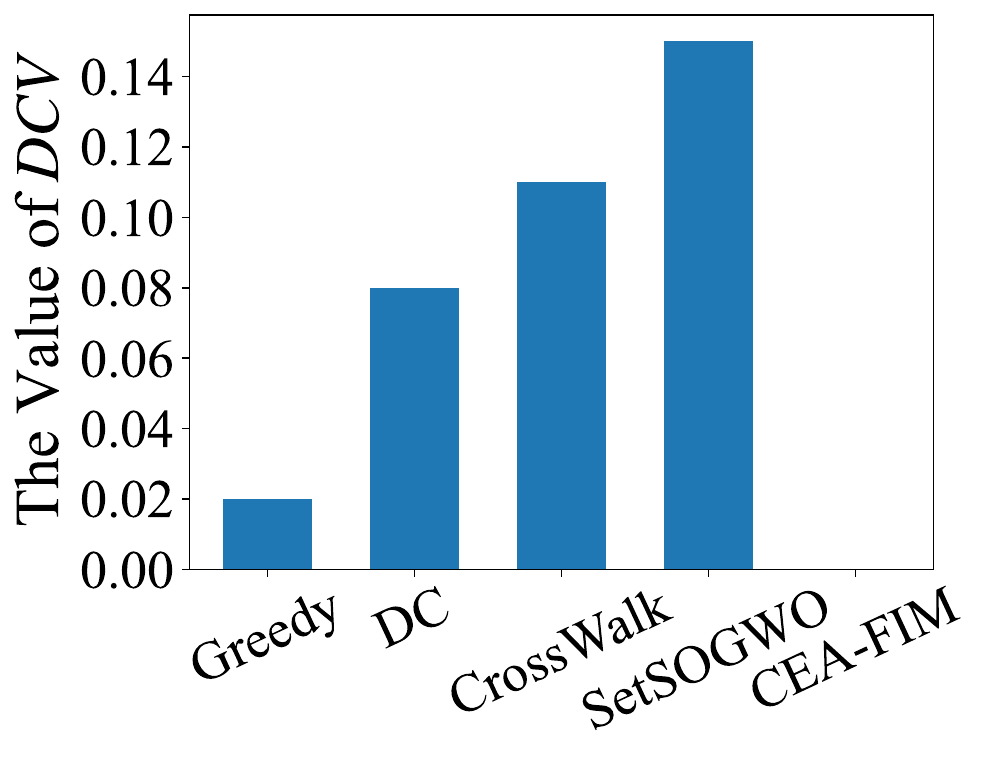}
			}
			\subfigure[]{
				\includegraphics[width=1.58in]{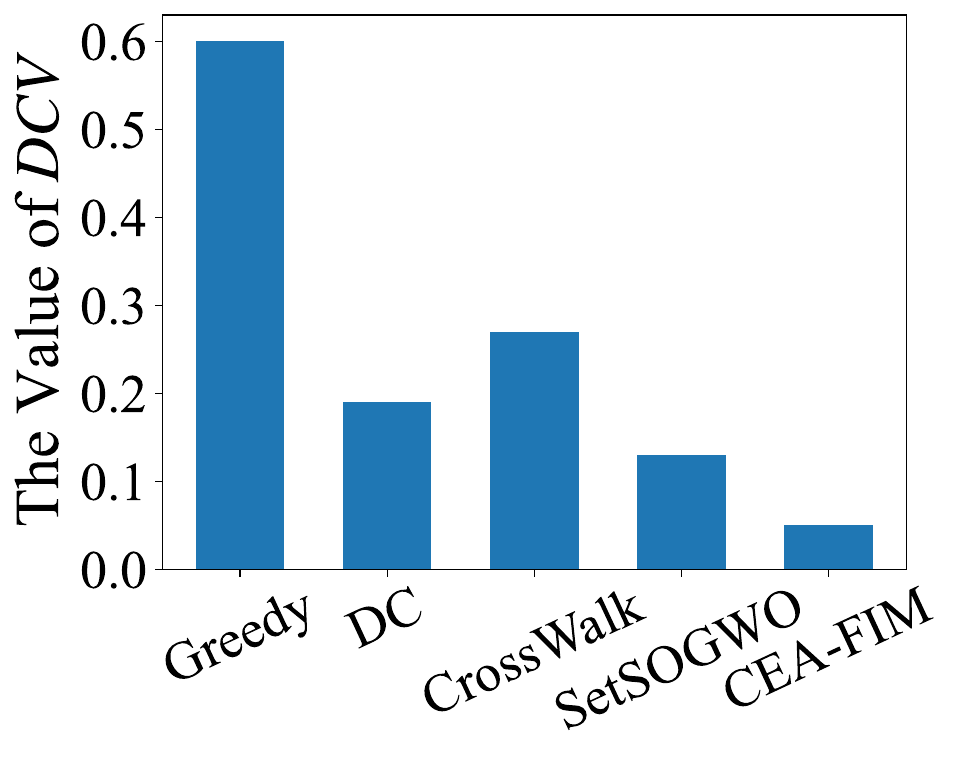}
			}
			\subfigure[]{
				\includegraphics[width=1.58in]{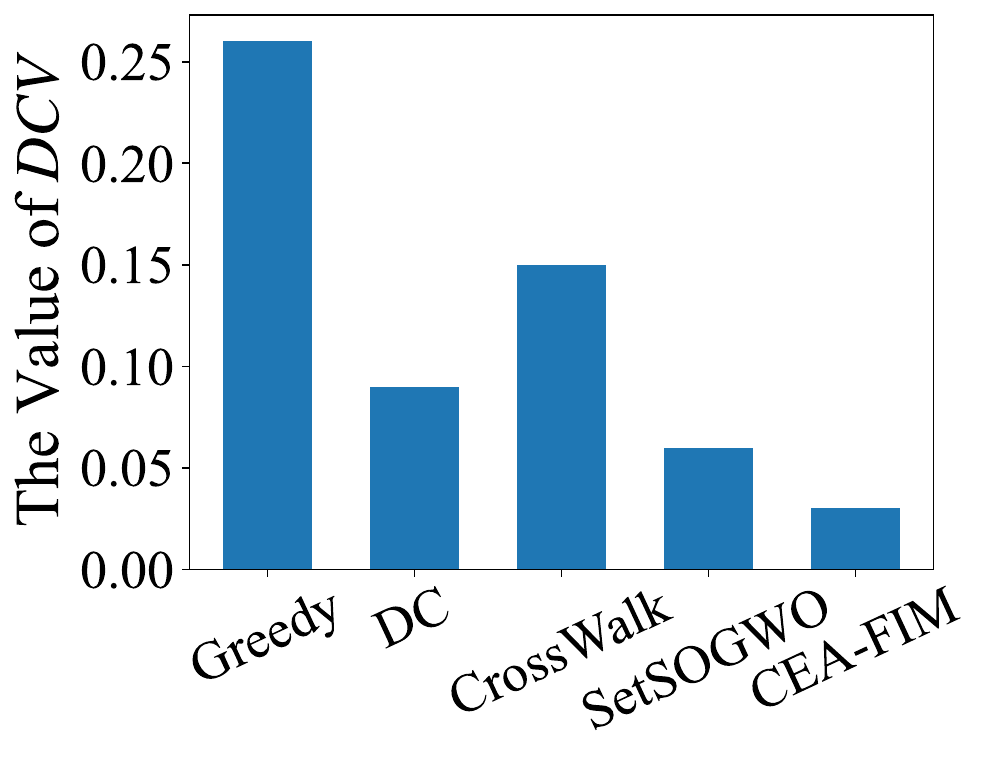}
			}
			\subfigure[]{
				\includegraphics[width=1.58in]{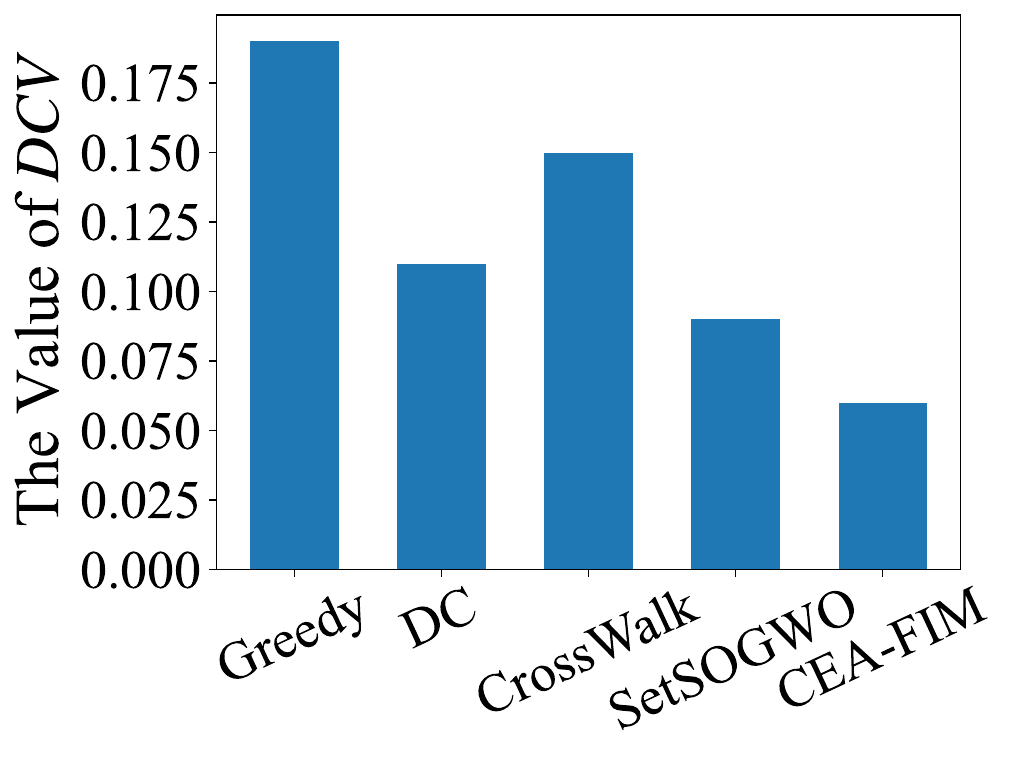}
			}
			\subfigure[]{
				\includegraphics[width=1.58in]{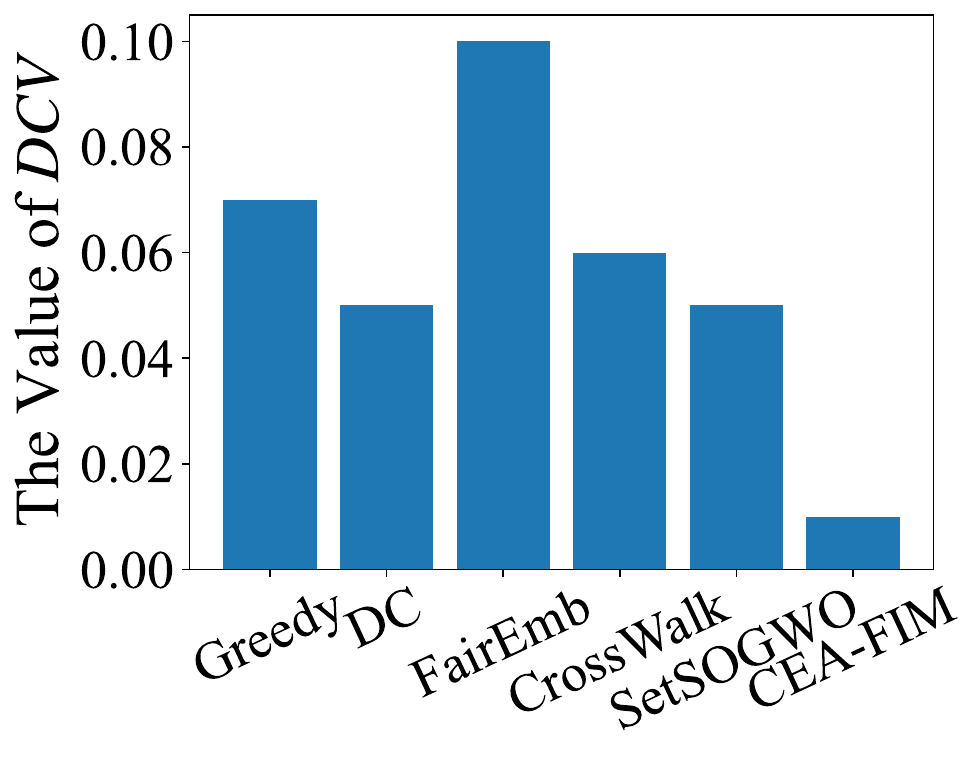}
			}
			\subfigure[]{
				\includegraphics[width=1.58in]{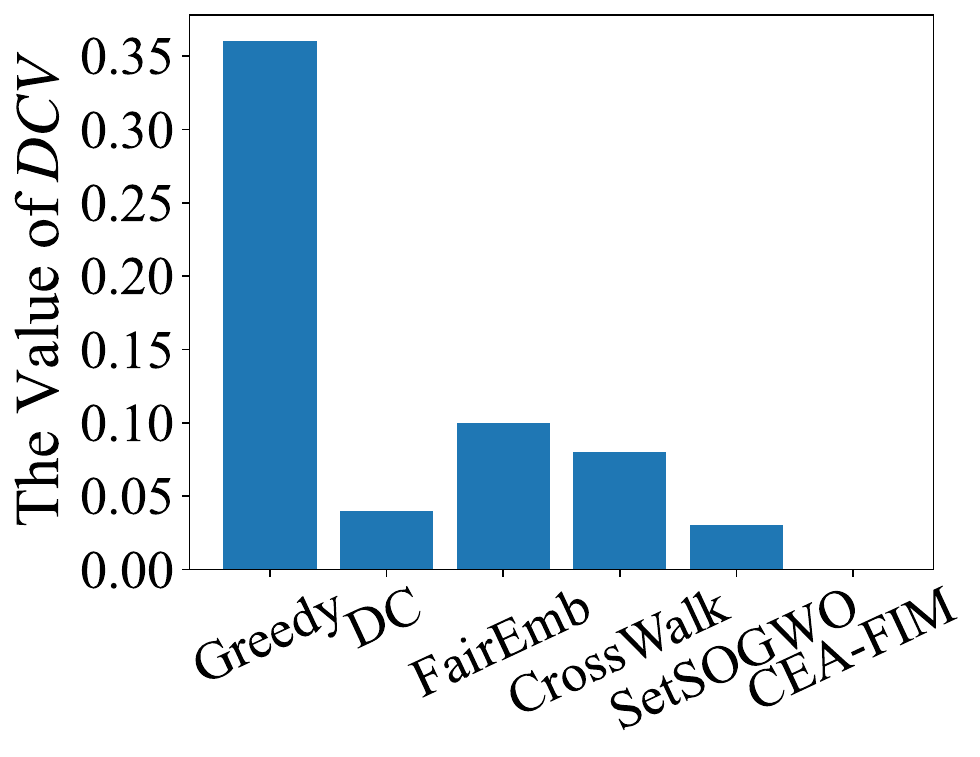}
			}
			\subfigure[]{
				\includegraphics[width=1.58in]{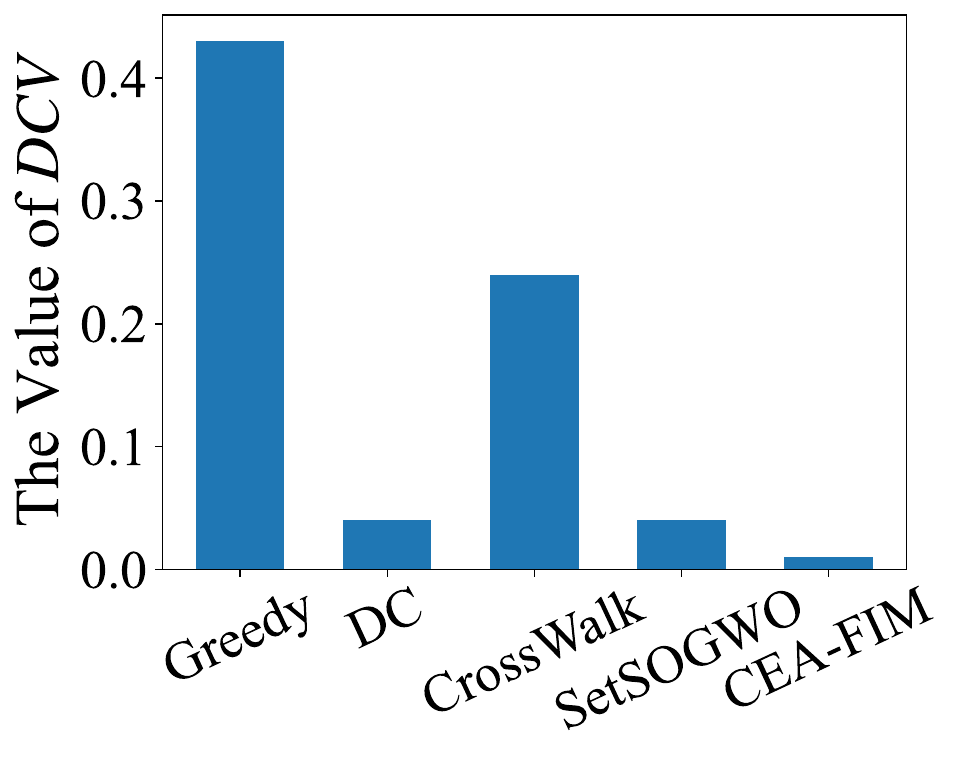}
			}
			\caption{\textcolor{black}{The comparison on the mean percentage violation of the constraints over all groups of different algorithms. The smaller values of $DCV$ are more desirable. (a) Rice-Facebook. (b) Twitter. (c) Synth1-region. (d) Synth1-ethnicity. (e) Synth1-age. (f) Synth1-gender. (g) Synth2. (h) Synth3.}}\label{fig:violation}\vspace{-0.2cm}
		\end{center}
	\end{figure}\vspace{-0.0cm}
	
	\begin{figure}[!htb] \vspace{-0.7cm}
		\begin{center}
			\centering
			\subfigure[]{
				\includegraphics[width=1.58in]{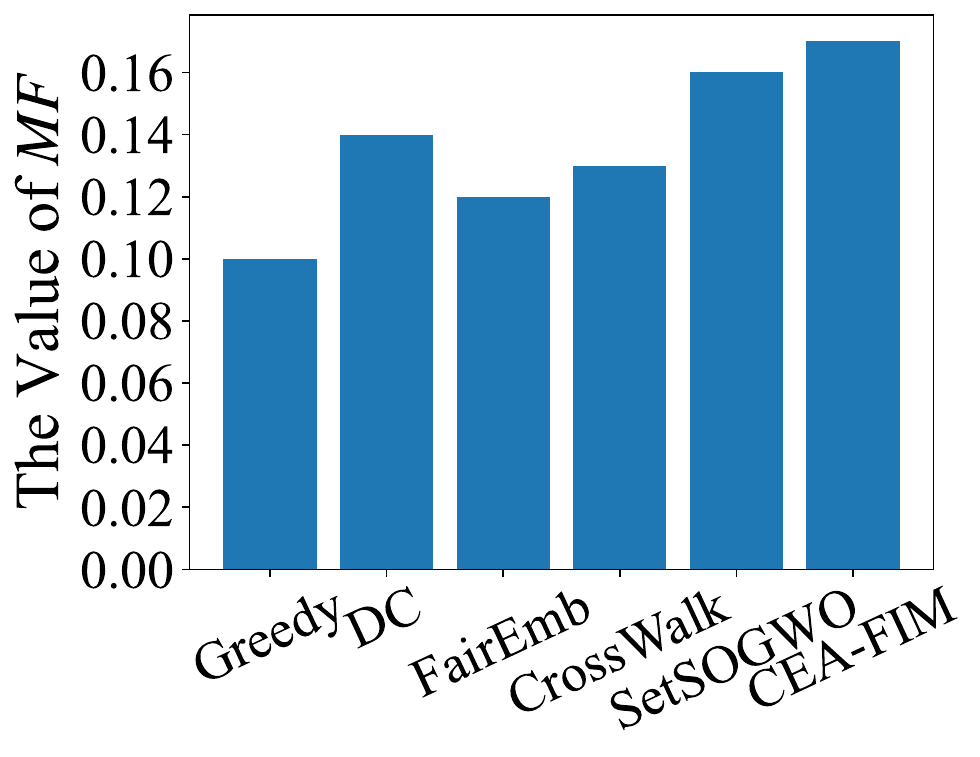}
			}
			\subfigure[]{
				\includegraphics[width=1.58in]{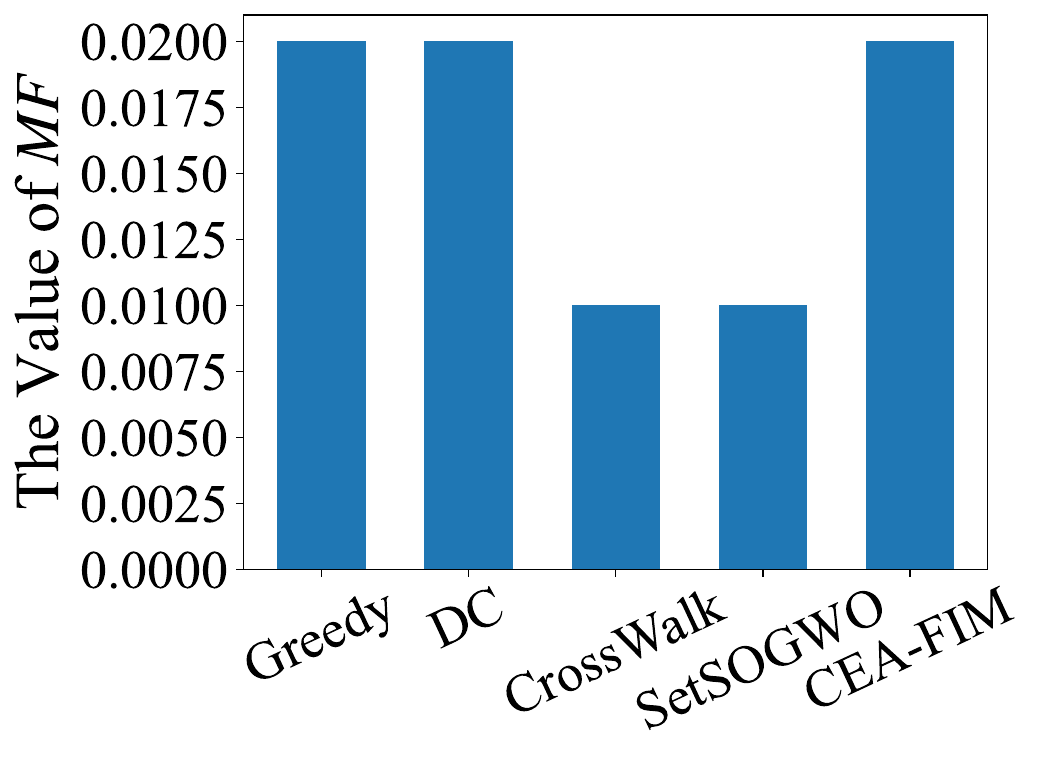}
			}
			\subfigure[]{
				\includegraphics[width=1.58in]{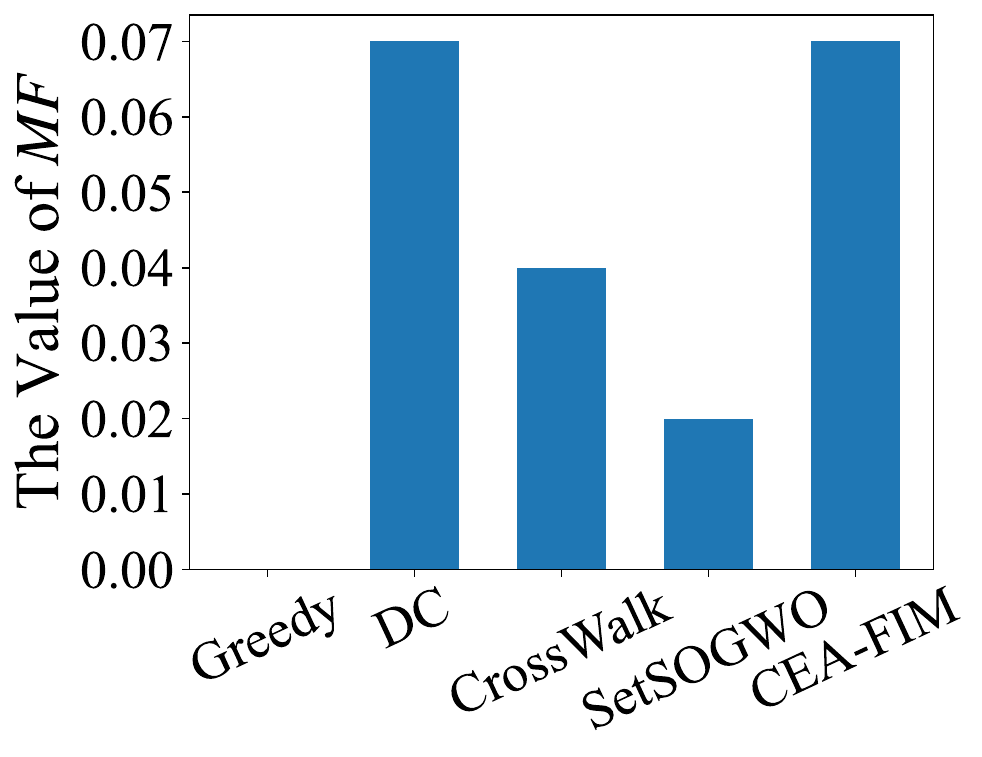}
			}
			\subfigure[]{
				\includegraphics[width=1.58in]{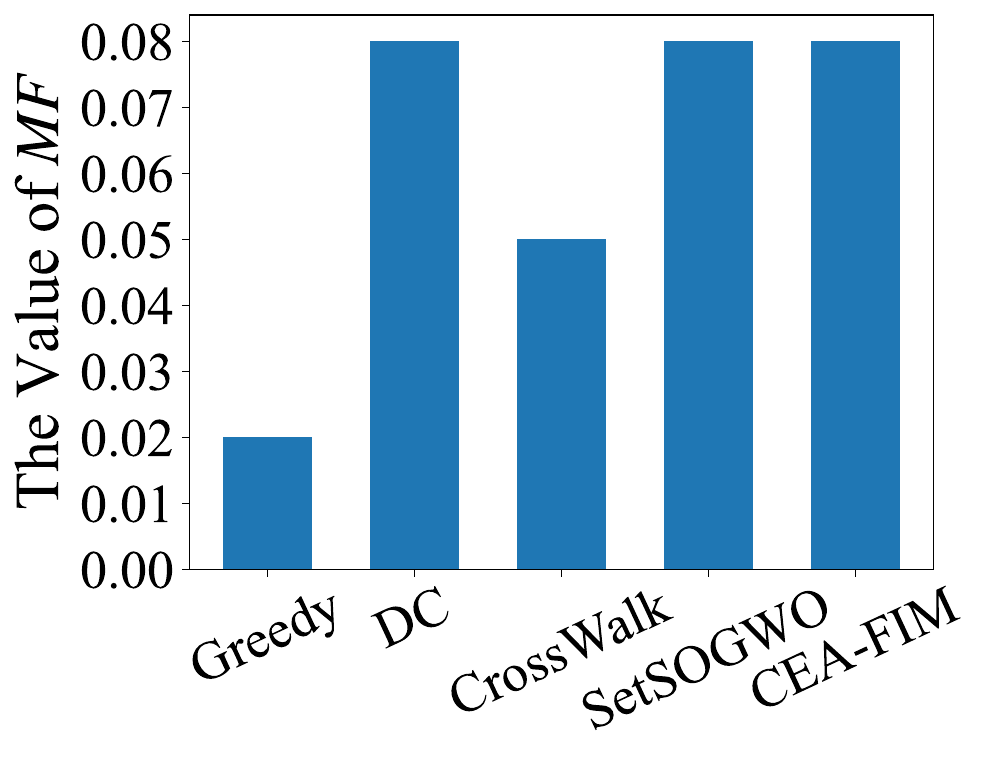}
			}
			\subfigure[]{
				\includegraphics[width=1.58in]{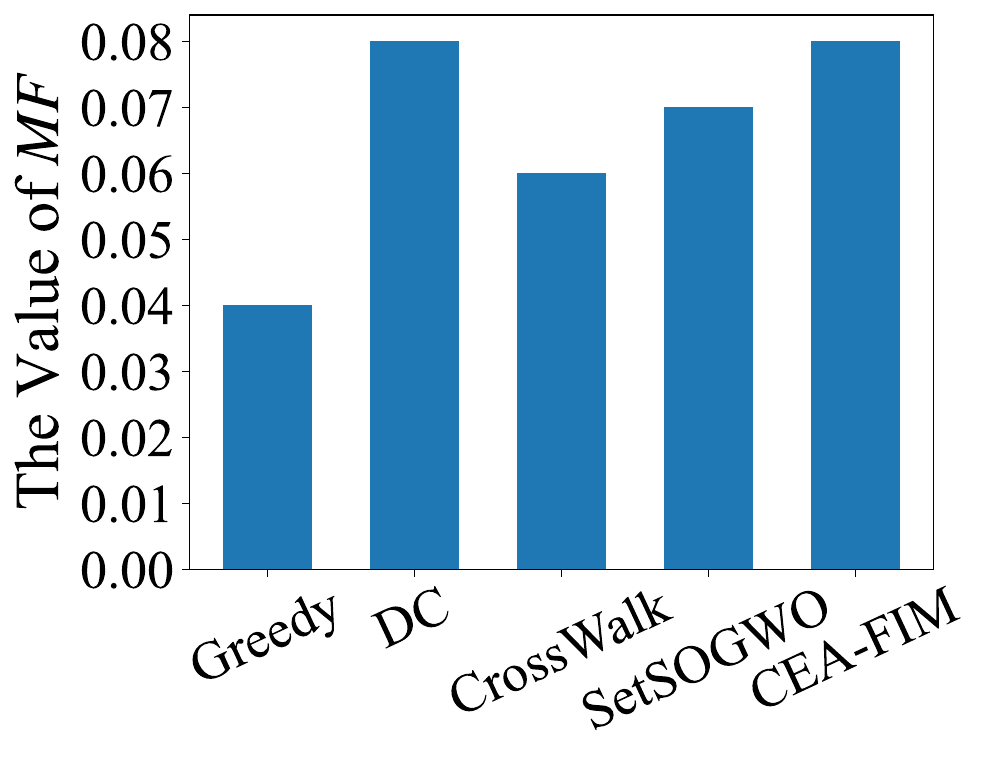}
			}
			\subfigure[]{
				\includegraphics[width=1.58in]{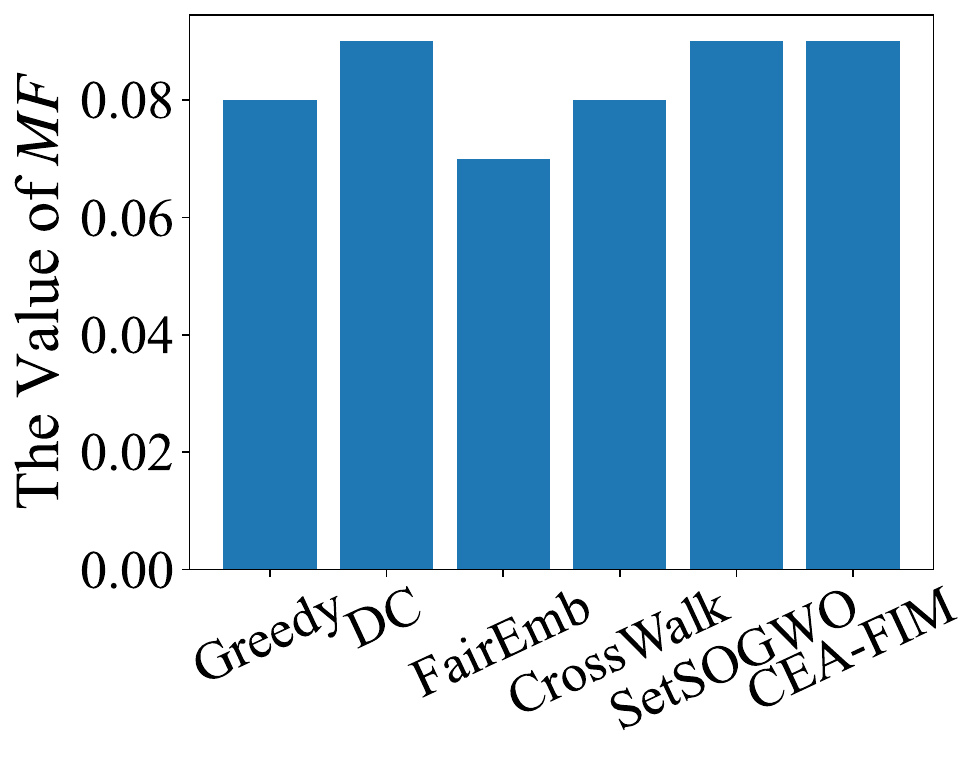}
			}
			\subfigure[]{
				\includegraphics[width=1.58in]{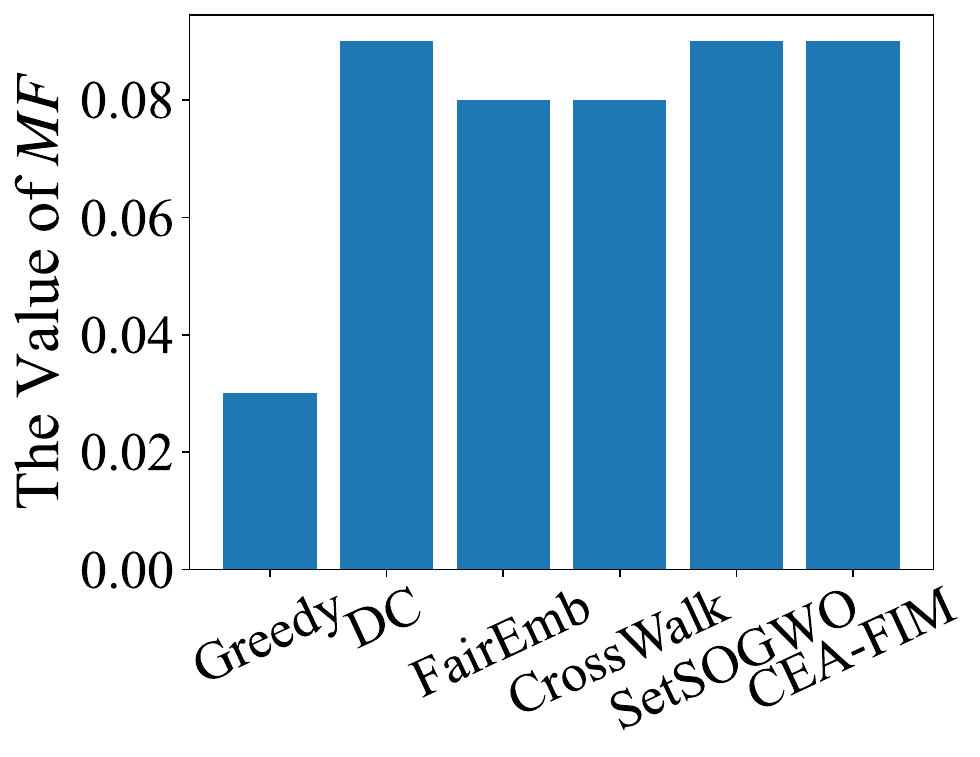}
			}
			\subfigure[]{
				\includegraphics[width=1.58in]{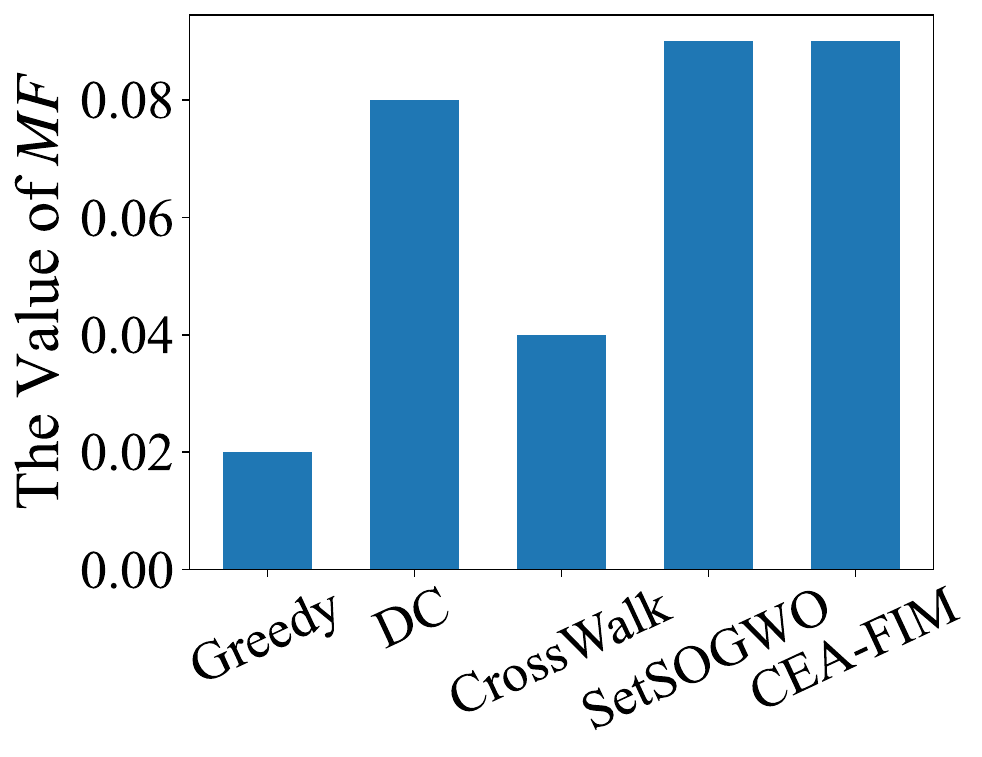}
			}
			\caption{\textcolor{black}{The comparison on the minimum fraction influenced over all groups of different algorithms. The larger values of $MF$ are more desirable. (a) Rice-Facebook. (b) Twitter. (c) Synth1-region. (d) Synth1-ethnicity. (e) Synth1-age. (f) Synth1-gender. (g) Synth2. (h) Synth3.}}\label{fig:min-fraction}\vspace{-0.2cm}
		\end{center}
	\end{figure}\vspace{-0.0cm}

	\begin{figure}[!htb] \vspace{-0.7cm}
		\begin{center}
			\centering
			\subfigure[]{
				\includegraphics[width=1.58in]{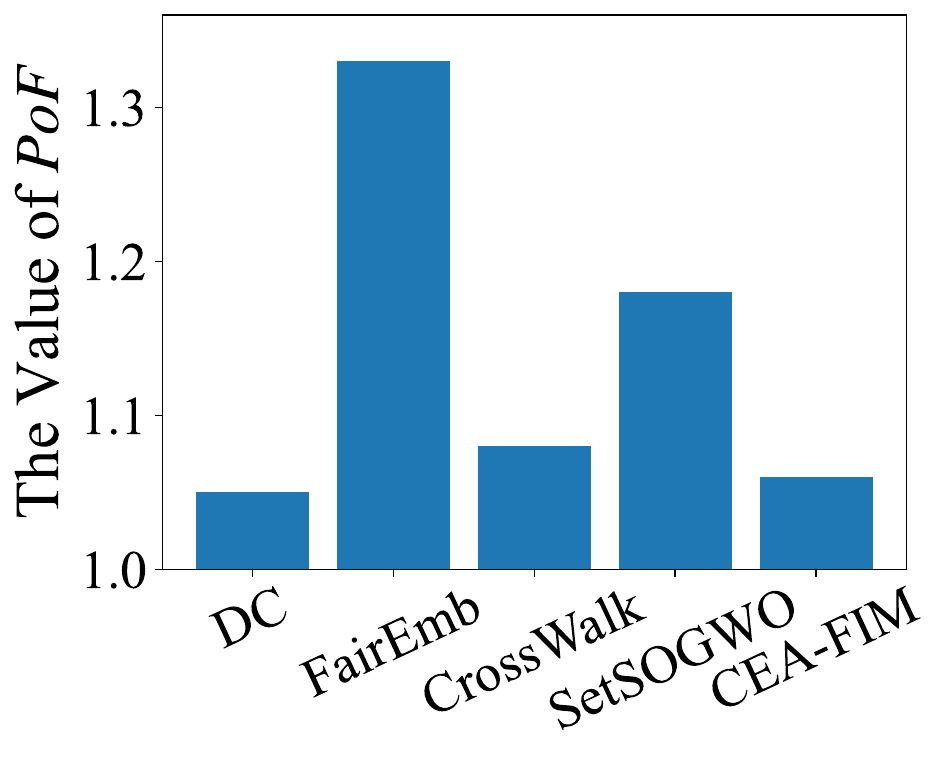}
			}
			\subfigure[]{
				\includegraphics[width=1.58in]{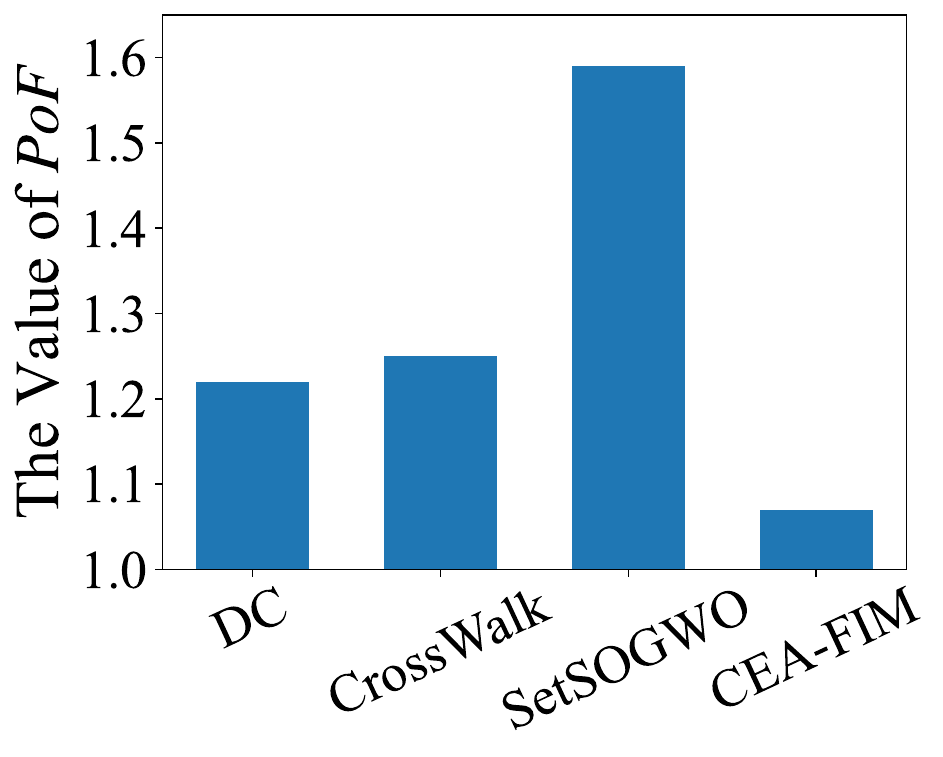}
			}
			\subfigure[]{
				\includegraphics[width=1.58in]{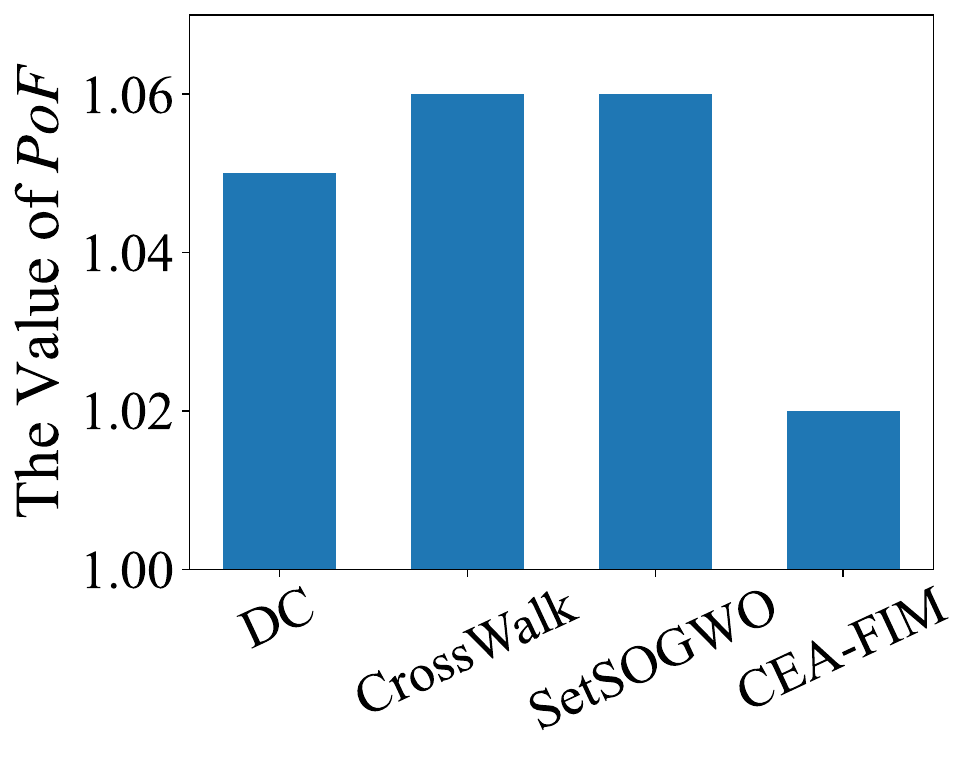}
			}
			\subfigure[]{
				\includegraphics[width=1.58in]{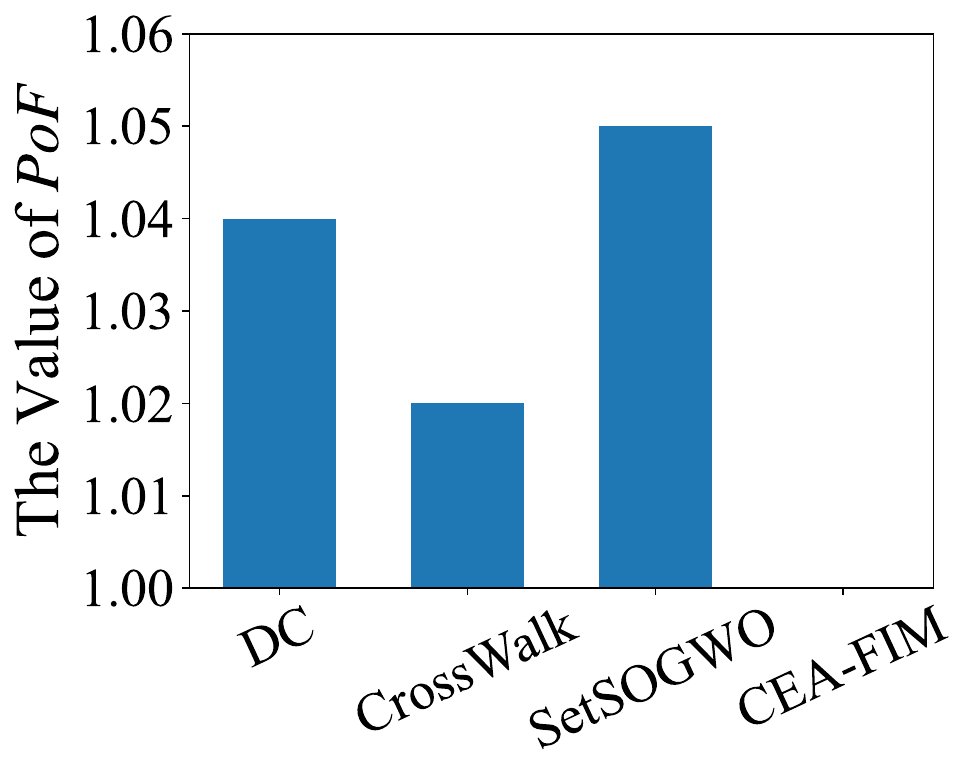}
			}
			\subfigure[]{
				\includegraphics[width=1.58in]{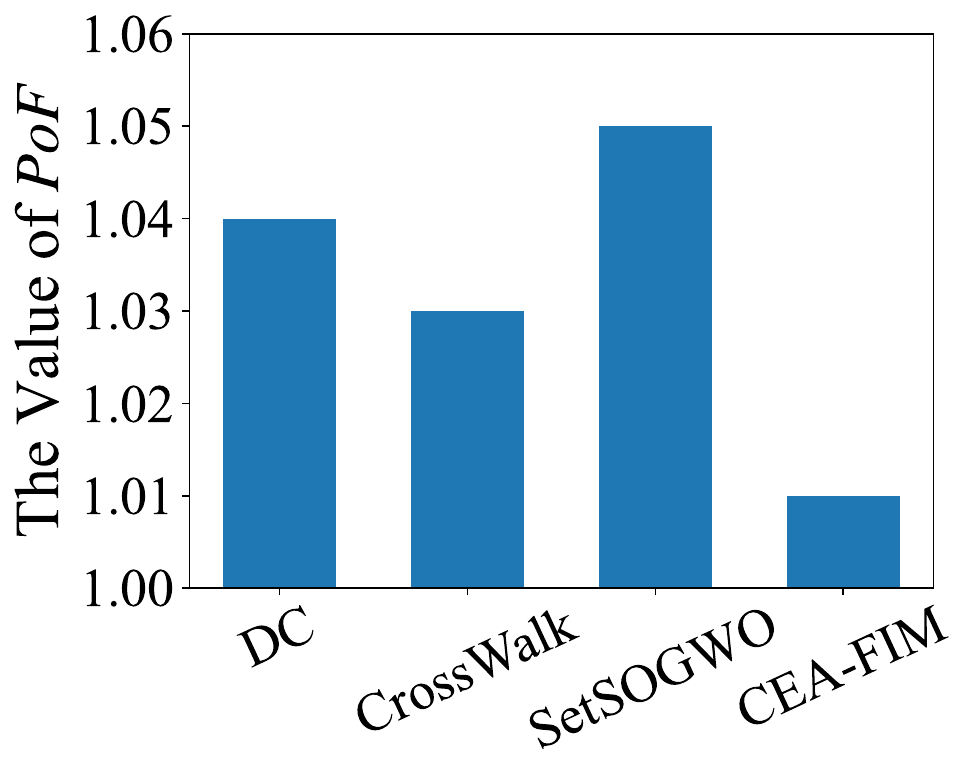}
			}
			\subfigure[]{
				\includegraphics[width=1.58in]{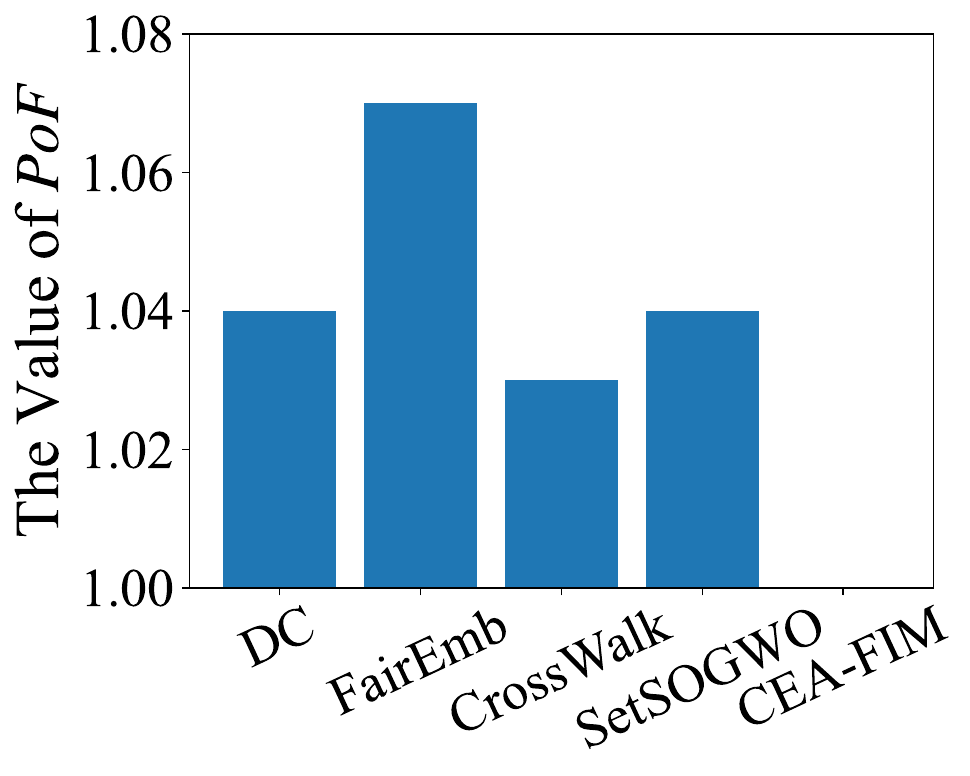}
			}
			\subfigure[]{
				\includegraphics[width=1.58in]{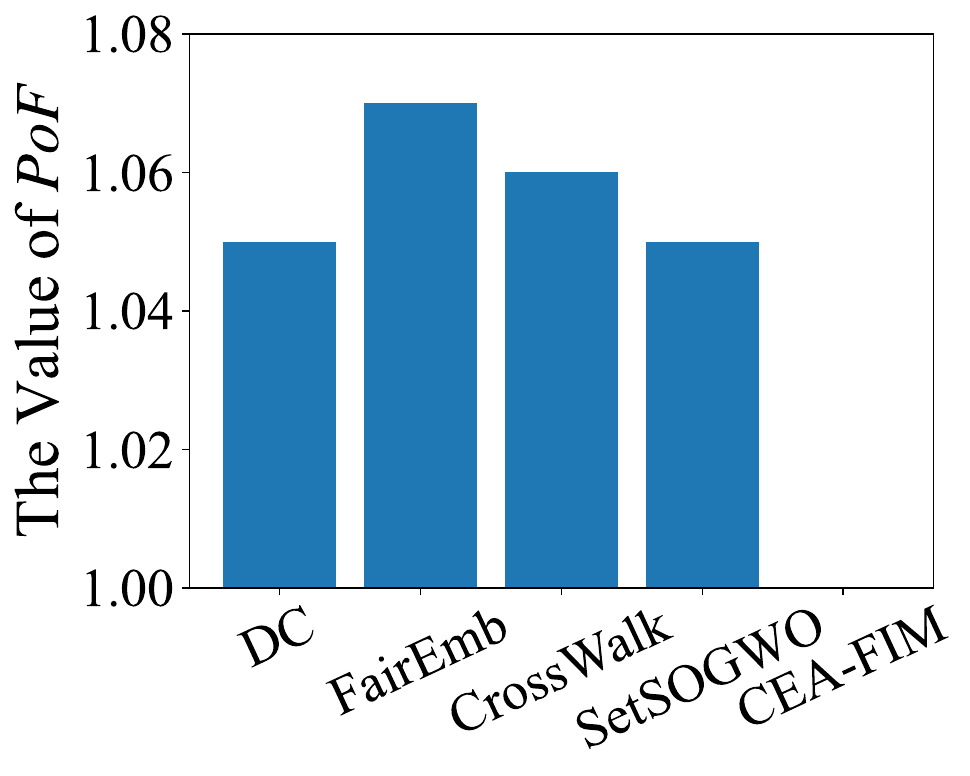}
			}
			\subfigure[]{
				\includegraphics[width=1.58in]{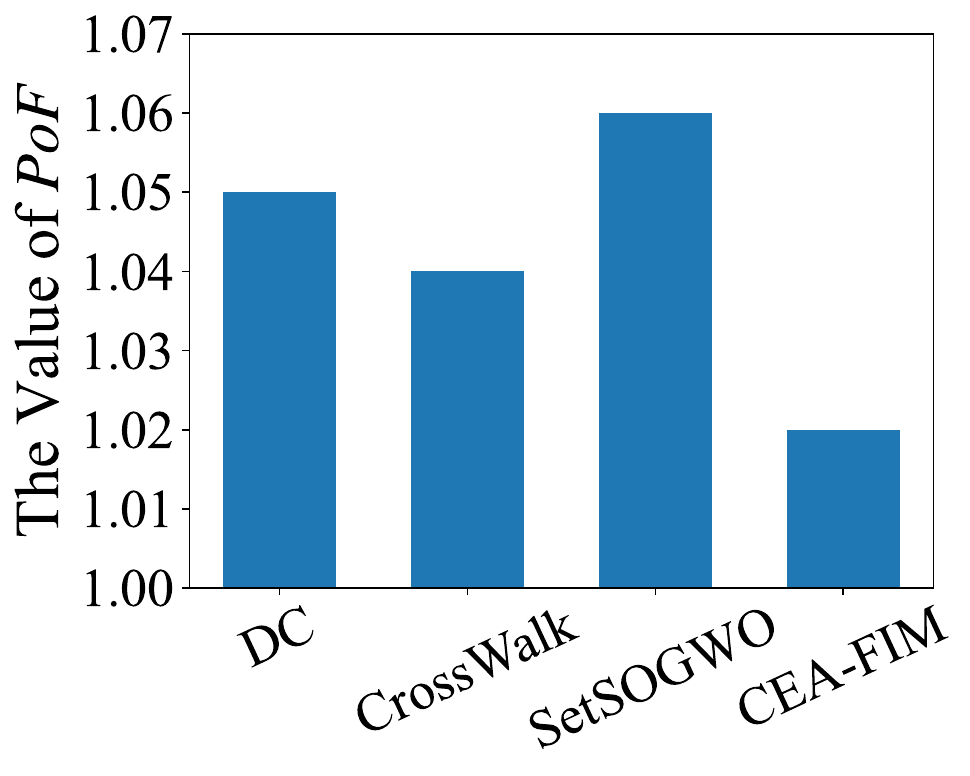}
			}
			\caption{\textcolor{black}{The comparison on the cost of achieving fairness of different algorithms. The smaller values of $PoF$ are more desirable. (a) Rice-Facebook. (b) Twitter. (c) Synth1-region. (d) Synth1-ethnicity. (e) Synth1-age. (f) Synth1-gender. (g) Synth2. (h) Synth3.}}\label{fig:pof}\vspace{-0.2cm}
		\end{center}
	\end{figure}\vspace{-0.0cm}

	\begin{table}[ht]
		\centering
		\begin{scriptsize}
			\caption{\textcolor{black}{The comparison on the evaluation function values of different algorithms.}}\label{tab:f-value} \vspace{-0.2cm}
			\setlength{\tabcolsep}{1mm}{
				\begin{tabular}{|c|c|c|c|c|c|c|}
					\hline
					Networks  &Greedy &DC &FairEmb &CrossWalk &\textcolor{black}{SetSOGWO} &CEA-FIM \\
					\hline
					Rice-Facebook & 0.01 & 0.14 & -0.01 & 0.12 &0.10 & \textbf{0.16}\\
					\hline
					Twitter & 0 & -0.06 & - & -0.10 &-0.15 & \textbf{0.02}\\
					\hline
					Synth1-region & -0.60 & -0.12 & - & 0.23 &-0.04 & \textbf{0.02}\\
					\hline
					Synth1-ethnicity & -0.24 & -0.01 & - & -0.10 &0.02 & \textbf{0.05}\\
					\hline
					Synth1-age & -0.15 & -0.03 & - & -0.09 &-0.02 & \textbf{0.02}\\
					\hline
					Synth1-gender & 0.01 & 0.04 & -0.03 & 0.02 &0.04 & \textbf{0.08}\\
					\hline
					Synth2 & -0.33 & 0.05 & -0.02 & 0 &0.06 & \textbf{0.09}\\
					\hline
					Synth3 & -0.41 & 0.04 & - & -0.20 &0.05 & \textbf{0.08}\\
					\hline
			\end{tabular}}
		\end{scriptsize}
	\end{table}\vspace{-0.0cm}
	
\begin{figure}[!htb]\vspace{-0.1cm}
		\small
		\begin{center}
			\centering
			\subfigure{
				\includegraphics[width=3.5in]{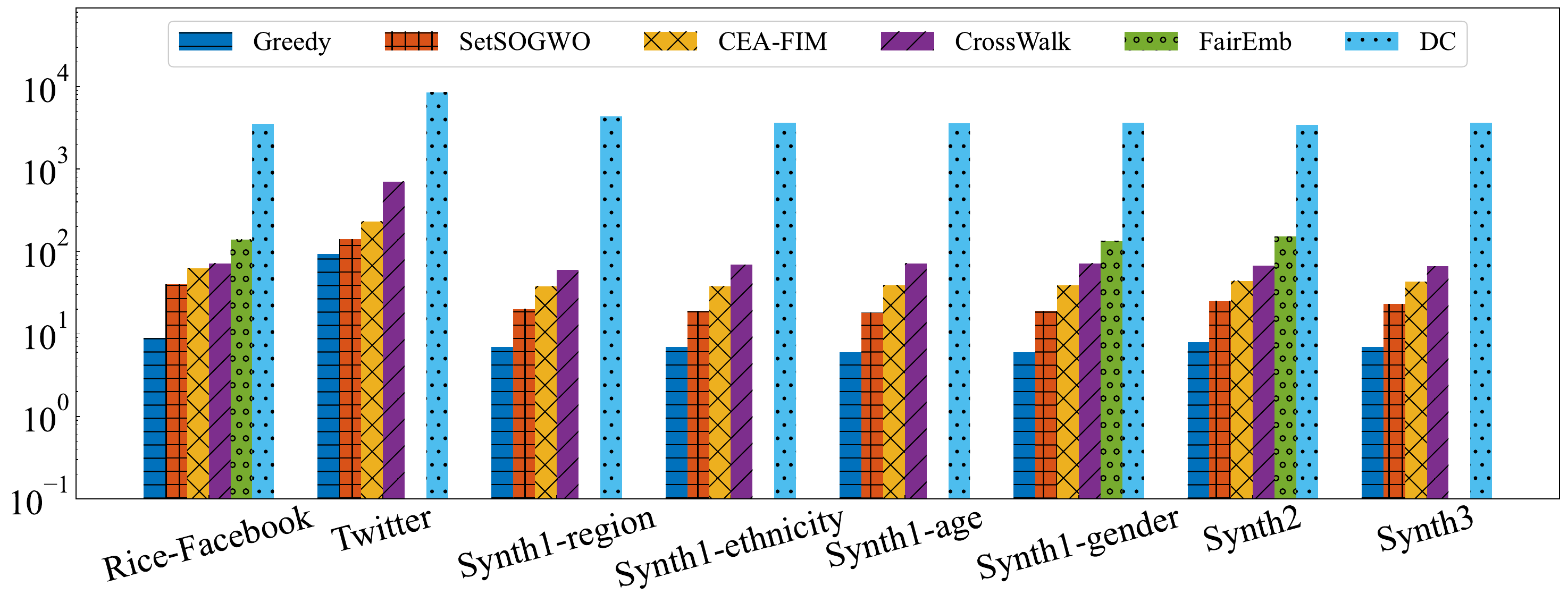}
			} \vspace{-0.5cm}
			\caption{\textcolor{black}{The running time of different algorithms on real-world and synthetic networks when $k=40$.}}\label{fig:time}
		\end{center}
	\end{figure}

	In order to validate the performance of our CEA-FIM algorithm, we compare the mean percentage of constraint violations (i.e., $DCV$ shown in Eq.~\eqref{eq:eq-dcv}) over all groups and the minimum fraction influenced (i.e., $MF$ shown in Eq.~\eqref{eq:eq-mf}) across all groups (see Figs.~\ref{fig:violation} and~\ref{fig:min-fraction}). In addition, in Table~\ref{tab:f-value} and Figs.~\ref{fig:pof} and~\ref{fig:time}, we report the cost of achieving fairness (i.e., $PoF$ shown in Eq.~\eqref{eq:eq-pof}), the value of the evaluation function (i.e., $F$ shown in Eq.~\eqref{eq:eq-MD}) and the running time of the \textcolor{black}{six} algorithms (i.e., Greedy, DC, FairEmb, CrossWalk, \textcolor{black}{SetSOGWO} and CEA-FIM). Note that we only report the performance of FairEmb on Rice-Facebook, Synth1-gender and Synth2, due to the fact that this baseline algorithm is merely applicable to networks of two groups.

	As is shown in Fig.~\ref{fig:violation}, Greedy generates the highest violations of diversity constraints in most cases. The reason why Greedy achieves such poor results is that it focuses on the influence spread without considering fairness. Furthermore, our proposed method CEA-FIM outperforms other algorithms in the vast majority of cases.  One exception is the mediocre performance of CEA-FIM on the network Rice-Facebook, while its performance on the minimum fraction influenced is the best. This may be due to the setting of the evaluation function, where $DCV$ and $MF$ are implicitly set the same weight (i.e., 0.5). It is worth noting that DC fails to obtain a good result on Twitter. This may be because its parameters are not suitable for this network, which is the largest network in the dataset. Nevertheless, the $DCV$ value obtained by our proposed algorithm CEA-FIM is zero, which means that each group in Twitter has received the resources it needs. Moreover, \textcolor{black}{SetSOGWO,} FairEmb and CrossWalk improve the fairness of the seed set $S$ to a certain extent, but there is still room for improvement. \textcolor{black}{In particular, SetSOGWO achieves the highest $DCV$ value on Twitter, which does not utilize the community structure information of social networks.} Fig.~\ref{fig:min-fraction} demonstrates that CEA-FIM outperforms other algorithms in terms of the minimum fraction influenced across all groups. Greedy achieves the smallest minimum fraction influenced in the majority of cases. Particularly, Greedy obtains a maximum value equal to zero on Synth1-region, implying that at least one group was not successfully influenced.
	
	In Fig.~\ref{fig:pof}, we report the cost of achieving fairness of these algorithms. It can be found that our proposed algorithm CEA-FIM outperforms other algorithms in most cases. The $PoF$ values of DC and CEA-FIM are very close. In addition, Table~\ref{tab:f-value} shows the evaluation function (i.e., $F$) values of these algorithms, where CEA-FIM achieves the highest value. Fig.~\ref{fig:time} shows the running time of \textcolor{black}{CEA-FIM and five comparison algorithms} with the seed set size of 40. The running time of Greedy is the shortest, which may be due to the fact that Greedy do not consider the fairness of information diffusion and the size of the adopted datasets is relative small. Note that the time complexity of Greedy is sensitive to the number of nodes in the network, for example, the performance gap between CEA-FIM and Greedy becomes smaller when they run on larger dataset Twitter. CEA-FIM, obviously, is nearly two orders of magnitude faster than DC, suggesting its applicability to large-scale social networks.
	
	\vspace{-0.0cm}

	In summary, the proposed method CEA-FIM achieves a better balance in terms of fairness and running time, which is mainly attributed the proposed community-based node selection strategy. We will verify it in the following subsection.

	\subsubsection{Effectiveness of Community-based Node Selection Strategy}\label{subsec:cns}
	\begin{table}[ht]
		\centering
		\begin{scriptsize}
			\caption{The comparison on the evaluation function values of REA-FIM and CEA-FIM.}\label{tab:rea-cea} \vspace{-0.2cm}
			\setlength{\tabcolsep}{6mm}{
			\begin{tabular}{|c|c|c|c|}
				\hline
				Networks  &REA-FIM &CEA-FIM \\
				\hline
				Rice-Facebook & 0.12 & \textbf{0.16}\\
				\hline
				Twitter & -0.12 & \textbf{0.02}\\
				\hline
				Synth1-region & 0.01 & \textbf{0.02}\\
				\hline
				Synth1-ethnicity & 0.03 & \textbf{0.05}\\
				\hline
				Synth1-age & -0.01 & \textbf{0.02}\\
				\hline
				Synth1-gender & 0.05 & \textbf{0.08}\\
				\hline
				Synth2 & 0.06 & \textbf{0.09}\\
				\hline
				Synth3 & 0.05 & \textbf{0.08}\\
				\hline
			\end{tabular}}
		\end{scriptsize}
	\end{table}\vspace{-0.0cm}
	
	To verify the effectiveness of the proposed community-based node selection strategy, we compare CEA-FIM with its variant REA-FIM on real-world and synthetic networks. Table~\ref{tab:rea-cea} shows the evaluation function (i.e., $F$) values of the two algorithms. From this table, we can observe that CEA-FIM obtains the best performance compared to REA-FIM. This result indicates that fairness can be better ensured by using community information to select nodes. Specifically, the advantage of CEA-FIM is most evident on Twitter. This is likely because there are many nodes in the network, and the random node selection strategy is hard to accurately select useful nodes. However, utilizing the information of the community structure makes it easy to find nodes that disseminate information fairly.

	\section{Conclusion And Future Work}\label{sec:conclude}
	
	In this paper, we proposed a community-based evolutionary algorithm called CEA-FIM for solving the FIM problem. We first modelled this problem as a single-objective optimization problem. Then, we designed an effective node selection strategy by utilizing the information of community structure. Last but not least, we proposed a community-based evolutionary algorithm to find a seed set to ensure the fair spread of information across all groups, including novel suggested initialization, crossover and mutation strategies. Also, we conducted extensive experiments on real-world and synthetic networks to validate the performance of CEA-FIM.
	
	The experimental results have shown the effectiveness of the proposed community-based evolutionary algorithm. In future research, we will conduct further research on the definition of fairness in FIM, which may have a great impact on minority groups. Besides, we plan to design efficient evolutionary algorithms to solve FIM problems under complex constraints, such as location, competition and time. Finally, we plan to apply CEA-FIM on dynamic and large-scale social networks, which are of great research value in real life.

\bibliographystyle{IEEEtran}
\bibliography{ref}

\appendix
\section{Supplementary Materials for CEA-FIM}

\subsection{\textcolor{black}{Computational Complexity of CEA-FIM}}
\textcolor{black}{According to the framework of the proposed algorithm CEA-FIM, there are three main factors affecting the efficiency of CEA-FIM, namely community detection, population initialization and population evolution. For the community detection phase, Louvain algorithm requires $\mathcal{O}(n\cdot \log(n))$ time. In the step of population initialization, the computation of node scores and the initialization of population can be completed in $\mathcal{O}(t(\varepsilon)\cdot n^2)$ and $\mathcal{O}(k\cdot n\cdot pop)$ time respectively in the worst case, where $t(\varepsilon)$ is the number of iterations associated with the threshold $\varepsilon$. In the last step, it takes $\mathcal{O}(k\cdot \delta \cdot n^2 \cdot w)$ time to prepare before using the evaluation function $F$ in the worst case, where $\delta$ is the number of live-edge graphs. Using quicksort, the sorting of individuals demands $\mathcal{O}(\delta \cdot n \cdot pop\cdot \log(pop))$ time. Both crossover and mutation operations need $\mathcal{O}(m\cdot k \cdot pop)$ time in the worst case. In order to select better individuals, it takes $\mathcal{O}(\delta\cdot n \cdot pop)$ time in the operation of selection. On the whole, the computational complexity of CEA-FIM scales as $\mathcal{O}(\delta \cdot n\cdot (k\cdot n\cdot w+pop \cdot \log(pop) \cdot g_{max}))$.}

\subsection{\textcolor{black}{Sensitivity of Parameter $\lambda$ in CEA-FIM}}\label{sec:parameter}
\begin{table*}[!t]
	\centering
	\caption{\textcolor{black}{The variations of $DCV$ with different $\lambda$ values when $k=40$.}}\label{tab:l-dcv}
	\setlength{\tabcolsep}{1.2mm}{
		\begin {tabular}{c c c c c c c c c c c c}
		\hline
		\diagbox{Networks}{$\lambda$} &0 &0.1 &0.2 &0.3 &0.4 &0.5 &0.6 &0.7 &0.8 &0.9 &1.0\\
		\hline
		\multirow{1}{*}{Rice-Facebook}
		&\bm{$0(1)$} &\bm{$0(1)$} &\bm{$0(1)$} &\bm{$0(1)$} &\bm{$0(1)$} &0.01(2) &0.03(3) &0.05(4) &0.05(4) &0.05(4) &0.06(5) \\
		\hline
		\multirow{1}{*}{Twitter}
		&\bm{$0(1)$} &\bm{$0(1)$} &\bm{$0(1)$} &\bm{$0(1)$} &\bm{$0(1)$} &\bm{$0(1)$} &\bm{$0(1)$} &\bm{$0(1)$} &\bm{$0(1)$} &\bm{$0(1)$} &0.03(2) \\
		\hline
		\multirow{1}{*}{Synth1-region}
		&\bm{$0.05(1)$} &\bm{$0.05(1)$} &\bm{$0.05(1)$} &\bm{$0.05(1)$} &\bm{$0.05(1)$} &\bm{$0.05(1)$} &\bm{$0.05(1)$} &\bm{$0.05(1)$} &\bm{$0.05(1)$} &0.07(2) &0.11(3) \\
		\hline
		\multirow{1}{*}{Synth1-ethnicity}
		&\bm{$0.03(1)$} &\bm{$0.03(1)$} &\bm{$0.03(1)$} &\bm{$0.03(1)$} &\bm{$0.03(1)$} &\bm{$0.03(1)$} &0.04(2) &\bm{$0.03(1)$} &0.04(2) &0.04(2) &0.06(3) \\
		\hline
		\multirow{1}{*}{Synth1-age}
		&\bm{$0.06(1)$} &\bm{$0.06(1)$} &\bm{$0.06(1)$} &\bm{$0.06(1)$} &\bm{$0.06(1)$} &\bm{$0.06(1)$} &0.07(2) &0.07(2) &0.07(2) &0.08(3) &0.09(4) \\
		\hline
		\multirow{1}{*}{Synth1-gender}
		&\bm{$0.01(1)$} &\bm{$0.01(1)$} &\bm{$0.01(1)$} &\bm{$0.01(1)$} &\bm{$0.01(1)$} &\bm{$0.01(1)$} &\bm{$0.01(1)$} &\bm{$0.01(1)$} &\bm{$0.01(1)$} &\bm{$0.01(1)$} &\bm{$0.01(1)$} \\
		\hline
		\multirow{1}{*}{Synth2}
		&\bm{$0(1)$} &\bm{$0(1)$} &\bm{$0(1)$} &\bm{$0(1)$} &\bm{$0(1)$} &\bm{$0(1)$} &\bm{$0(1)$} &0.01(2) &0.02(3) &0.02(3) &0.02(3) \\
		\hline
		\multirow{1}{*}{Synth3}
		&\bm{$0.01(1)$} &\bm{$0.01(1)$} &\bm{$0.01(1)$} &\bm{$0.01(1)$} &\bm{$0.01(1)$} &\bm{$0.01(1)$} &0.02(2) &0.02(2) &\bm{$0.01(1)$} &0.02(2) &0.03(3) \\
		\hline
		\multicolumn{1}{c}{$Ave\_Ranking$} &\textbf{1.0} &\textbf{1.0} &\textbf{1.0} &\textbf{1.0} &\textbf{1.0} &1.1 &1.6 &1.8 &1.9 &2.3 &3.0 \\
		\hline
\end{tabular}}
\end{table*}
\begin{table*}[!t]
\centering
\caption{\textcolor{black}{The variations of $MF$ with different $\lambda$ values when $k=40$.}}\label{tab:l-mf}
\setlength{\tabcolsep}{1.2mm}{
	\begin {tabular}{c c c c c c c c c c c c}
	\hline
	\diagbox{Networks}{$\lambda$} &0 &0.1 &0.2 &0.3 &0.4 &0.5 &0.6 &0.7 &0.8 &0.9 &1.0\\
	\hline
	\multirow{1}{*}{Rice-Facebook}
	&0.13(5) &0.15(4) &0.15(4) &0.15(4) &0.16(3) &0.17(2) &\bm{$0.19(1)$} &\bm{$0.19(1)$} &\bm{$0.19(1)$} &\bm{$0.19(1)$} &\bm{$0.19(1)$} \\
	\hline
	\multirow{1}{*}{Twitter}
	&\bm{$0.02(1)$} &\bm{$0.02(1)$} &\bm{$0.02(1)$} &\bm{$0.02(1)$} &\bm{$0.02(1)$} &\bm{$0.02(1)$} &\bm{$0.02(1)$} &\bm{$0.02(1)$} &\bm{$0.02(1)$} &\bm{$0.02(1)$} &\bm{$0.02(1)$} \\
	\hline
	\multirow{1}{*}{Synth1-region}
	&0.06(2) &0.06(2) &\bm{$0.07(1)$} &\bm{$0.07(1)$} &\bm{$0.07(1)$} &\bm{$0.07(1)$} &\bm{$0.07(1)$} &\bm{$0.07(1)$} &\bm{$0.07(1)$} &\bm{$0.07(1)$} &\bm{$0.07(1)$} \\
	\hline
	\multirow{1}{*}{Synth1-ethnicity}
	&\bm{$0.08(1)$} &\bm{$0.08(1)$} &\bm{$0.08(1)$} &\bm{$0.08(1)$} &\bm{$0.08(1)$} &\bm{$0.08(1)$} &\bm{$0.08(1)$} &\bm{$0.08(1)$} &\bm{$0.08(1)$} &\bm{$0.08(1)$} &\bm{$0.08(1)$} \\
	\hline
	\multirow{1}{*}{Synth1-age}
	&0.07(2) &0.07(2) &0.07(2) &0.07(2) &0.07(2) &\bm{$0.08(1)$} &\bm{$0.08(1)$} &\bm{$0.08(1)$} &\bm{$0.08(1)$} &\bm{$0.08(1)$} &\bm{$0.08(1)$} \\
	\hline
	\multirow{1}{*}{Synth1-gender}
	&\bm{$0.09(1)$} &\bm{$0.09(1)$} &\bm{$0.09(1)$} &\bm{$0.09(1)$} &\bm{$0.09(1)$} &\bm{$0.09(1)$} &\bm{$0.09(1)$} &\bm{$0.09(1)$} &\bm{$0.09(1)$} &\bm{$0.09(1)$} &\bm{$0.09(1)$} \\
	\hline
	\multirow{1}{*}{Synth2}
	&0.09(2) &0.09(2) &0.09(2) &0.09(2) &0.09(2) &0.09(2) &0.09(2) &0.09(2) &\bm{$0.10(1)$} &\bm{$0.10(1)$} &\bm{$0.10(1)$} \\
	\hline
	\multirow{1}{*}{Synth3}
	&\bm{$0.09(1)$} &\bm{$0.09(1)$} &\bm{$0.09(1)$} &\bm{$0.09(1)$} &\bm{$0.09(1)$} &\bm{$0.09(1)$} &\bm{$0.09(1)$} &\bm{$0.09(1)$} &\bm{$0.09(1)$} &\bm{$0.09(1)$} &\bm{$0.09(1)$} \\
	\hline
	\multicolumn{1}{c}{$Ave\_Ranking$} &1.9 &1.8 &1.6 &1.6 &1.5 &1.3 &1.1 &1.1 &\textbf{1.0} &\textbf{1.0} &\textbf{1.0} \\
	\hline
\end{tabular}}
\end{table*}
\begin{table*}[!t]
\centering
\caption{\textcolor{black}{The variations of $PoF$ with different $\lambda$ values when $k=40$.}}\label{tab:l-pof}
\setlength{\tabcolsep}{1.2mm}{
\begin {tabular}{c c c c c c c c c c c c}
\hline
\diagbox{Networks}{$\lambda$} &0 &0.1 &0.2 &0.3 &0.4 &0.5 &0.6 &0.7 &0.8 &0.9 &1.0\\
\hline
\multirow{1}{*}{Rice-Facebook}
&\bm{$1.04(1)$} &1.07(3) &1.06(2) &1.06(2) &1.06(2) &1.06(2) &1.09(4) &1.10(5) &1.10(5) &1.11(6) &1.12(7) \\
\hline
\multirow{1}{*}{Twitter}
&1.10(4) &1.08(2) &1.09(3) &1.09(3) &1.09(3) &\bm{$1.07(1)$} &1.08(2) &\bm{$1.07(1)$} &1.08(2) &1.10(4) &1.11(5) \\
\hline
\multirow{1}{*}{Synth1-region}
&1.03(2) &1.03(2) &1.03(2) &1.03(2) &1.03(2) &\bm{$1.02(1)$} &\bm{$1.02(1)$} &1.03(2) &1.03(2) &1.03(2) &1.04(3) \\
\hline
\multirow{1}{*}{Synth1-ethnicity}
&1.01(2) &\bm{$1.00(1)$} &1.01(2) &\bm{$1.00(1)$} &1.01(2) &\bm{$1.00(1)$} &1.01(2) &1.01(2) &1.01(2) &1.01(2) &1.01(2) \\
\hline
\multirow{1}{*}{Synth1-age}
&1.02(2) &1.02(2) &1.02(2) &\bm{$1.01(1)$} &\bm{$1.01(1)$} &\bm{$1.01(1)$} &\bm{$1.01(1)$} &\bm{$1.01(1)$} &1.02(2) &1.02(2) &1.02(2) \\
\hline
\multirow{1}{*}{Synth1-gender}
&\bm{$1.00(1)$} &\bm{$1.00(1)$} &\bm{$1.00(1)$} &\bm{$1.00(1)$} &\bm{$1.00(1)$} &\bm{$1.00(1)$} &\bm{$1.00(1)$} &\bm{$1.00(1)$} &\bm{$1.00(1)$} &\bm{$1.00(1)$} &\bm{$1.00(1)$} \\
\hline
\multirow{1}{*}{Synth2}
&\bm{$1.00(1)$} &1.01(2) &\bm{$1.00(1)$} &1.01(2) &\bm{$1.00(1)$} &\bm{$1.00(1)$} &\bm{$1.00(1)$} &1.01(2) &1.01(2) &1.01(2) &1.01(2) \\
\hline
\multirow{1}{*}{Synth3}
&\bm{$1.02(1)$} &\bm{$1.02(1)$} &\bm{$1.02(1)$} &\bm{$1.02(1)$} &1.03(2) &\bm{$1.02(1)$} &\bm{$1.02(1)$} &\bm{$1.02(1)$} &1.03(2) &1.03(2) &1.03(2) \\
\hline
\multicolumn{1}{c}{$Ave\_Ranking$} &1.8 &1.8 &1.8 &1.6 &1.8 &\textbf{1.1} &1.6 &1.9 &2.3 &2.6 &3.0 \\
\hline
\end{tabular}}
\end{table*}

\textcolor{black}{In the proposed algorithm CEA-FIM, a weight parameter $\lambda$ is introduced in order to design an appropriate evaluation function. To determine the best value of $\lambda$, we run the proposed algorithm on all networks with different $\lambda$ values (i.e., $\lambda$ increases from 0 to 1.0 at the step of 0.1). In Tables~\ref{tab:l-dcv},~\ref{tab:l-mf} and~\ref{tab:l-pof}, the rankings of different weight parameters $\lambda$ of CEA-FIM in terms of $DCV$, $MF$, and $PoF$ are shown in brackets, respectively. As shown in Tables~\ref{tab:l-dcv} and~\ref{tab:l-mf}, the value of $DCV$ decreases with the decrease of $\lambda$ and the value of $MF$ increases with the increase of $\lambda$. When $\lambda=0.5$, CEA-FIM achieves good tradeoff on both $DCV$ and $MF$. Furthermore, as shown in Table~\ref{tab:l-pof}, the value of $PoF$ decreases with fluctuations as $\lambda$ increases in the interval [0, 0.5] and increases as $\lambda$ increases in the interval [0.5, 1]. The value of $PoF$ ranks lowest when $\lambda$ is equal to 0.5. Therefore, considering the fairness and influence spread of the seed set, $\lambda$ is set to 0.5 in this paper.}

\end{document}